\documentclass{pasj01}
\usepackage{graphicx}
\usepackage{natbib}

\def\FeXXV{{Fe}\emissiontype{XXV}}
\def\kms{\rm km\:s^{-1}}

\newcommand{\aav}[1]{\left\langle{#1}\right\rangle}

\begin{document}

\title{Constraining hydrostatic mass bias of galaxy clusters with high-resolution X-ray spectroscopy}

\author{Naomi \textsc{Ota}\altaffilmark{1}}
\altaffiltext{1}{Department of Physics, Nara Women's University, Kitauoyanishi-machi, Nara, Nara 630-8506, Japan}
\email{naomi@cc.nara-wu.ac.jp}

\author{Daisuke \textsc{Nagai}\altaffilmark{2,3}}
\email{daisuke.nagai@yale.edu}

\author{Erwin T. \textsc{Lau}\altaffilmark{2,3}}
\email{erwin.lau@yale.edu}

\altaffiltext{2}{Department of Physics, Yale University, New Haven, CT 06520, USA}
\altaffiltext{3}{Yale Center for Astronomy and Astrophysics, Yale University, New Haven, CT 06520, USA}

\maketitle

\begin{abstract}
Gas motions in galaxy clusters play important roles in determining the properties of the intracluster medium (ICM) and in the constraint of cosmological parameters via X-ray and Sunyaev-Zel'dovich effect observations of galaxy clusters. The Hitomi measurements of gas motions in the core of the Perseus Cluster have provided insights into the physics in galaxy clusters. The XARM mission, equipped with the Resolve X-ray micro-calorimeter, will continue Hitomi's legacy by measuring ICM motions through Doppler shifting and broadening of emission lines in a larger number of galaxy clusters, and at larger radii. In this work, we investigate how well we can measure bulk and turbulent gas motions in the ICM with XARM, by analyzing mock XARM simulations of galaxy clusters extracted from cosmological hydrodynamic simulations.  We assess how photon counts, spectral fitting methods, multiphase ICM structure, deprojections, and region selection affect the measurements of gas motions. We first show that XARM is capable of recovering the underlying spherically averaged turbulent and bulk velocity profiles for dynamically relaxed clusters to within $\sim 50\%$ with a reasonable amount of photon counts in the X-ray emission lines. We also find that there are considerable azimuthal variations in the ICM velocities, where the velocities measured in a single azimuthal direction can significantly deviate from the true value even in dynamically relaxed systems. Such variation must be taken into account when interpreting data and developing observing strategies. We will discuss the prospect of using the upcoming XARM mission to measure non-thermal pressure and to correct for the hydrostatic mass bias of galaxy clusters. Our results are broadly applicable for future X-ray missions, such as Athena and Lynx.
\end{abstract}

%-------------------------------------------------%
\section{Introduction}
%-------------------------------------------------%

In the hierarchical structure formation scenario, galaxy clusters are believed to grow through mergers and continuous mass accretion through the cosmic-web of large-scale structures. These mergers and accretion events stir up the hot gas in the deep gravitational potential well of the cluster, generating large-scale motions \citep{vazzaetal09, vazzaetal11, miniati14,miniati15}. Non-thermal pressure provided by these internal gas motions provides an extra non-thermal pressure support to the intracluster medium (ICM) \citep{lauetal09, battagliaetal12, nelsonetal14b, shietal14, shietal16}, introduces bias in the hydrostatic mass estimates of galaxy clusters \citep[e.g.,][]{rasiaetal06, nagaietal07b, nelsonetal12, ettorietal13}, and serves as one of the primary sources of systematic uncertainties in cluster-based cosmological constraints \citep{allenetal08, vikhlininetal09, planckXXIV16}. While there are indirect constraints on the level of gas motions through spatial fluctuations in thermal pressure \citep{schueckeretal04} and X-ray surface brightness \citep{churazovetal12,walkeretal15}, direct observational constraints from emission lines were missing, and only upper limits were available before the launch the Hitomi satellite (\citealp[and references therein]{otaetal07b, ota12}; \citealp[but see also]{sandersetal13, tamuraetal14} \citealp{liuetal15,otaetal16}).

The Hitomi X-ray observatory \citep{takahashietal16,takahashietal14}, equipped with high energy resolution ($\sim 5$~eV) Soft X-ray Spectrometer (SXS) \citep{kelleyetal16, mitsudaetal14}, has made the first direct measurement of bulk and turbulent gas motions, through shifting and broadening of the 6.7~keV \FeXXV~K$\alpha$ line \citep{hitomi16, hitomi17a}.  The Hitomi result is also consistent with the inferred levels of gas motions from Chandra analysis of X-ray surface brightness fluctuations \citep{zhuravlevaetal17} as well as predictions of hydrodynamical cosmological simulations that include the effects of both AGN (active galactic nuclei)  bubbles and cluster mergers and accretion \citep{lauetal17, bournesijacki17}.  

The X-ray Astronomy Recovery Mission (XARM) will continue the legacy of the short-live Hitomi mission. Its X-ray micro-calorimeter, Resolve is identical to the Soft X-ray Spectrometer (SXS) onboard Hitomi. It will measure turbulent and bulk motions in the ICM for a larger sample of clusters, and at larger cluster-centric radii beyond the cluster core. Such measurements will, for example, enable us to identify the role of mergers and accretion in driving gas motions in cluster cores and outer regions, as well as to constrain the level of non-thermal pressure and the hydrostatic mass bias of galaxy clusters. 
There are several questions that must be addressed before we can analyze and interpret the XARM data in a meaningful way.  How well can XARM measure and separate bulk and turbulent motions in clusters? What are optimal observing and analysis strategies? How can we use XARM to measure the non-thermal pressure in the ICM and correct for the hydrostatic mass bias of galaxy clusters? 

The primary goal of this work is to address these questions by analyzing mock XARM observations of galaxy clusters extracted from high-resolution hydrodynamical cosmological simulations. These simulations capture internal motions in the ICM generated by mergers and accretion events in the hierarchical structure formation model. We create a realistic mock XARM data of simulated clusters, analyze them using the data analysis pipeline used to analyze real X-ray data, and compare derived velocity measurements to ``true'' values measured directly in the simulations. We find that the spherically mass-weighted averaged ICM temperature and velocity profiles can be measured with the accuracy of better than 50\%, which leads to good recovery of the hydrostatic mass bias to 5\%, provided that there are at least 200 photons in the 6.7~keV \FeXXV~K$\alpha$ line.  However, our results also reveal considerable azimuthal variations in both bulk and turbulent velocities in the ICM (by up to factors of 2) at a given radius even for dynamically relaxed looking clusters.  We discuss implications of these azimuthal variations in correcting the hydrostatic mass bias using XARM cluster observations. 

This paper is organized as follows. Section~\ref{sec:sim} describes numerical simulations and mock XARM simulations used in this work.  Results are presented in section~\ref{sec:results}. We present discussion and conclusions in section~\ref{sec:discussion} and \ref{sec:conclusions}, respectively.

%-------------------------------------------------%
\section{Simulations}
\label{sec:sim}
%-------------------------------------------------%

%-------------------------------------------------%
\subsection{Hydrodynamical simulations}
%-------------------------------------------------%

In this paper, we used galaxy clusters selected from high-resolution cosmological hydrodynamical simulations from \citet{nagaietal07a} and \citet{nagaietal07b} (hereafter N07), performed with non-radiative gas physics. These simulations are performed using the Adaptive Refinement Tree (ART) $N$-body$+$gas-dynamics code \citep{kra99,kra02,ruddetal08}, an Eulerian code that uses adaptive refinement in space and time, and non-adaptive refinement in mass \citep{klypinetal01}, allowing us to achieve the dynamic ranges needed resolve high density regions in the cluster within the cosmological volume. Being an Eulerian mesh-based code with low numerical viscosity, the ART code is able to capture contact discontinuities, shocks, and turbulence, making it particularly ideal for theoretical and computational modeling of gas motions in galaxy clusters.  

Likely targets for XARM would be nearby X-ray luminous massive clusters. Thus, we focus on two X-ray luminous clusters, CL104 and CL101, at $z=0$, with the core-excised X-ray temperature of $T_{\rm X}=7.7$ and $8.7$~keV, respectively.  Each cluster is simulated using a $128^3$ uniform grid with 8 levels of refinement. Clusters are selected from 120$h^{-1}$~Mpc computational boxes, achieving peak spatial resolution of $\approx 3.6h^{-1}$~kpc, sufficient to resolve dense gas clumps in the ICM. The dark matter particle mass in the region surrounding the cluster is $9\times 10^8 h^{-1}M_{\odot}$, while the rest of the box volume is simulated with lower mass and spatial resolution. We refer readers to N07 for further details. 

\begin{figure}[t]
\begin{center}
\includegraphics[width=7.5cm]{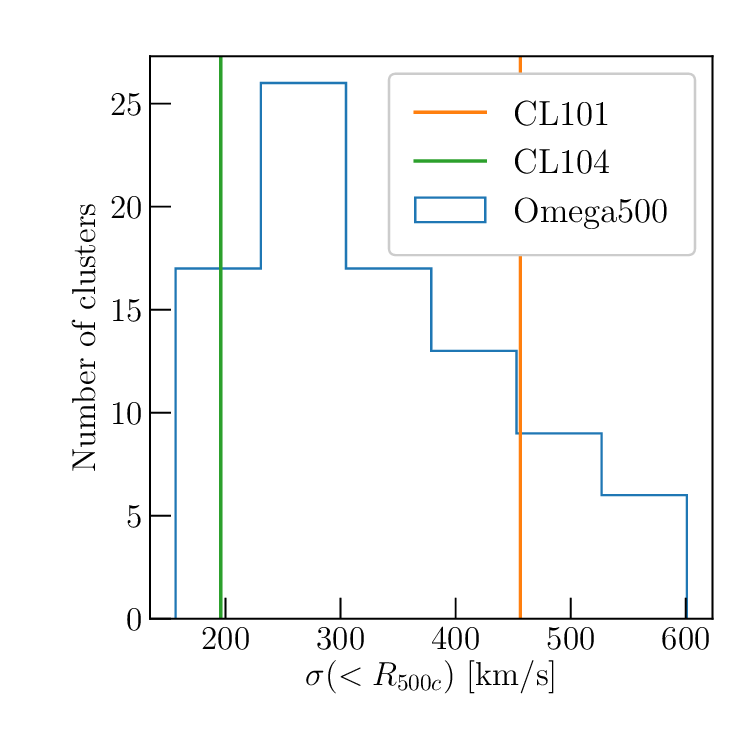}
\end{center}
\caption{Velocity dispersion of gas measured within $r_{500c}$ of our simulated clusters. The green and orange lines represent the relaxed and merging simulated clusters, CL104 and CL101, respectively. The blue line represents the distribution of the turbulent motions for the mass-limited sample of clusters from the Omega500 cosmological simulation. Note that CL104 and CL101 occupy the two ends of the distribution.}
\label{fig:sample}
\end{figure}

To investigate the dependence of the dynamical state of the cluster on ICM velocity measurements, these two clusters are selected based in their apparent relaxation state. CL104 is a massive, relaxed cluster that has not undergone any major merger for the past 8~Gyrs. CL101 is a similarly massive cluster, but it has undergone recent major mergers  (at $z\sim 0.1$ and $z\sim 0.25$), leaving two subclusters in its core that have been identified by visual inspection and are masked out for analysis. Figure~\ref{fig:sample} compares the velocity dispersion of gas, $\sigma$, within $r_{500c}$ for these two clusters to those measured for a larger, mass-limited sample of simulated galaxy clusters from the Omega500 cosmological simulations with non-radiative gas physics \citep{nelsonetal14a}, demonstrating that CL104 and CL101 represent the upper and lower ends of the level of gas motions, respectively. 

In our previous work \citep{nagaietal13}, we investigated effects of baryonic physics on the velocity structure of the ICM, by analyzing CL104 and CL101 performed with varying gas physics. The first set is performed with non-radiative gas physics. The second set includes radiative cooling, star formation, metal enrichment, and stellar feedback (CSF). The third set includes CSF and energy feedback from AGN (CSF+AGN). By comparing these three runs, we found that, while the baryonic physics can have significant impact on the properties of cluster cores, the velocity structure outside the cluster core ($r\gtrsim 0.15 r_{500}$) is relatively unaffected by the baryonic physics. Since the present work concerns the measurements of large-scale gas motions outside of cluster cores regions, we focus our analyses on the non-radiative simulations and discuss their limitations and caveats in section~\ref{sec:discussion}.

Note that these simulations follow the mass assembly process, such as mergers and accretion, of galaxy clusters in concordance with the $\Lambda$CDM model, and they are capable of reproducing the observed thermodynamic profiles of the X-ray emitting gas at the level of $10\%$ outside of cluster cores \citep{nagaietal07a}. Moreover, these simulations predict a significant level of the non-thermal pressure due to gas motions \citep{lauetal09,nelsonetal14b}, which introduces the biases in the hydrostatic mass estimates \citep{nagaietal07b,nelsonetal14a} as well as the observed power spectrum of the thermal Sunyaev-Zel'dovich effect \citep{shawetal10}. Therefore, these simulations are capable of robust modeling of gas motions generated by cosmic accretion outside of the cluster cores, and upcoming XARM measurements will provide the first observational tests of the level of gas motions predicted by our cluster simulations. 

%%%%%%%%%%%%%%%%%%%%%%%%%%%%%%%%%%%%%%%%%%%%%%%%%%%%%%%
%%%%%%%%%%%%%%%%%%%%%%%%%%%%%%%%%%%%%%%%%%%%%%%%%%%%%%%

%%%%%%%%%%%%%%%%%%%%%%%%%%%%%%%%%%%%%%%%%%%%%%%%%%%%%%%
\begin{figure}[t]
\begin{center}
\includegraphics[width=7.5cm]{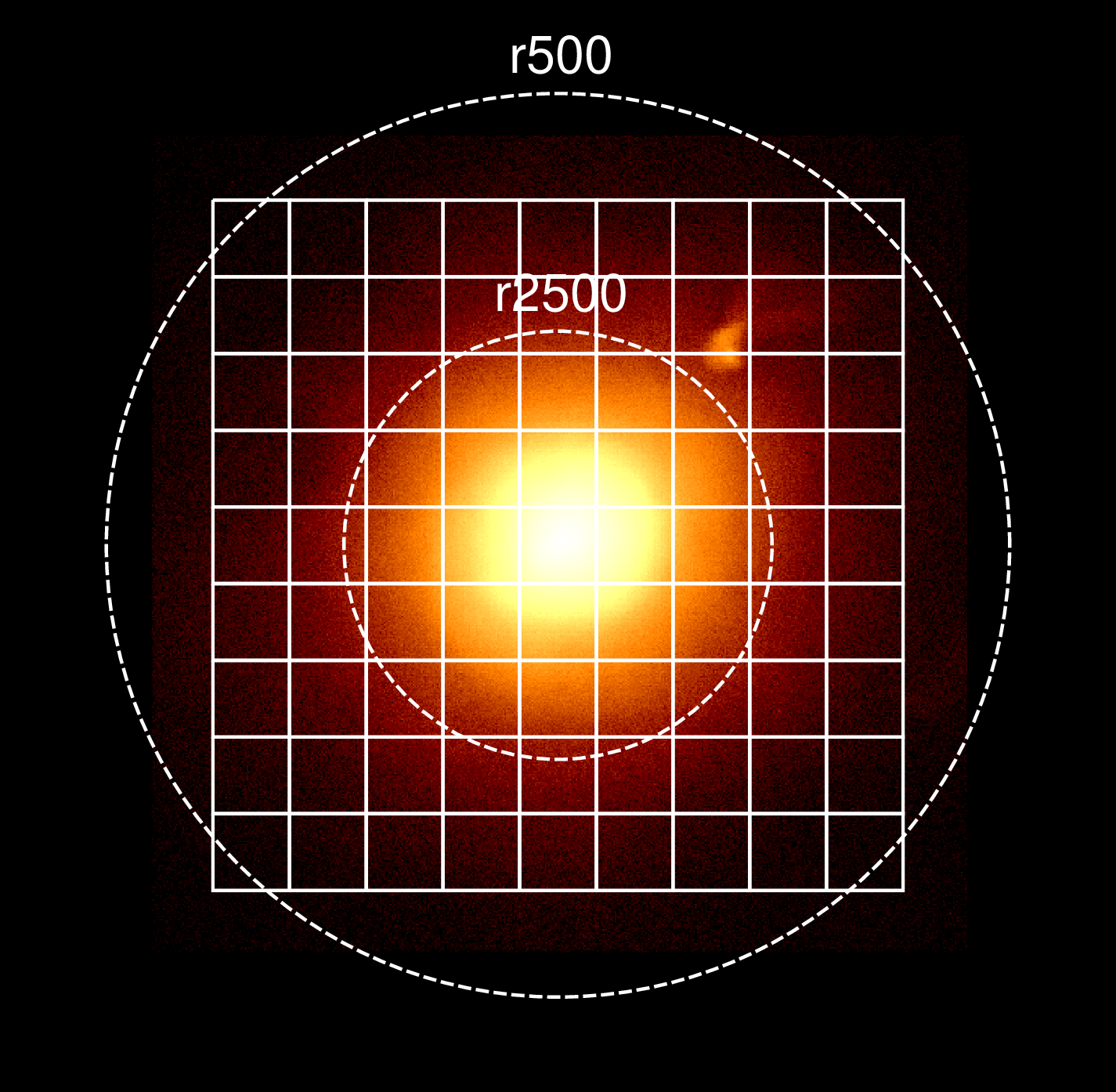}
\end{center}
\caption{Mock XARM image of CL104 in the $[0.3,10.0]$~keV band. $9\times 9$ boxes, each with a size of $3\arcmin\times3\arcmin$, were used for the analysis of projected velocity field. The grid $(i,j)$ at the bottom-left corner is $(1,1)$, and the grid label $i/j$ indicates the cell located at the upward/rightward direction from the origin, respectively.
\label{fig:CL104_x_box_region}}

\end{figure}
%%%%%%%%%%%%%%%%%%%%%%%%%%%%%%%%%%%%%%%%%%%%%%%%%%%%%%%

%%%%%%%%%%%%%%%%%%%%%%%%%%%%%%%%%%%%%%%%%%%%%%%%%%%%%%%
\begin{figure}
\begin{center}
\includegraphics[width=8.5cm]{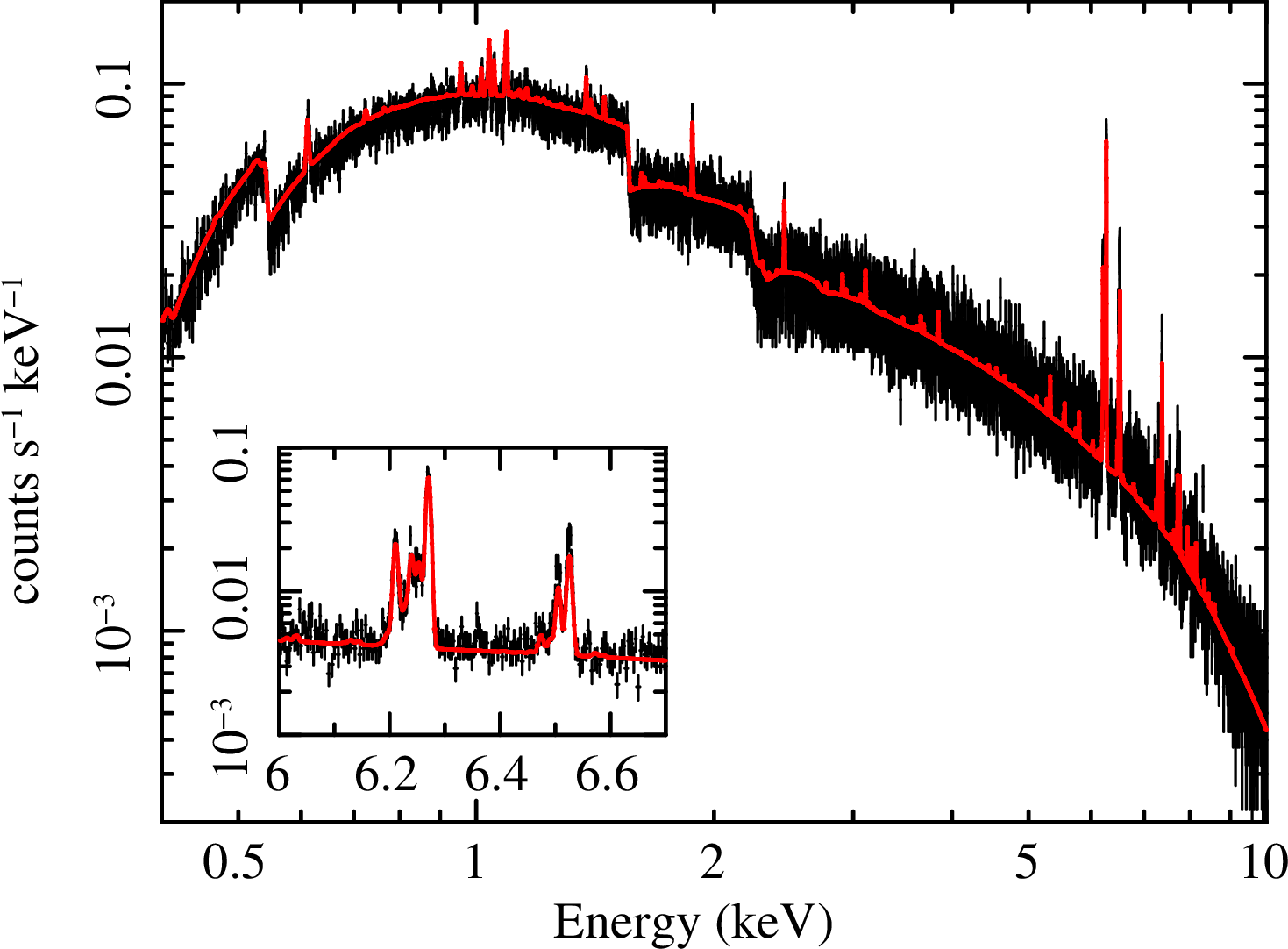}
\end{center}
\caption{Mock XARM spectrum for the energy range $[0.3,10.0]$~keV extracted from the grid $(i,j)=(7,5)$ defined in figure~\ref{fig:CL104_x_box_region} of CL104 with total exposure time of 2~Ms. The inset shows the spectrum around the \FeXXV~K$\alpha$ line (redshifted to $6.27$~keV) where the velocity constraints come from. The red line shows the best model BAPEC fit of the spectrum. The data are grouped by a factor of 8 and 1 spectral bin = 8~eV in the plot. 
\label{fig:spec_CL104_x_2e6_cir1_3}}
\end{figure}
%%%%%%%%%%%%%%%%%%%%%%%%%%%%%%%%%%%%%%%%%%%%%%%%%%%%%%%

%-------------------------------------------------%
\subsection{Mock XARM simulations}
%-------------------------------------------------%

For each simulated cluster, we generate a mock XARM observations for three orthogonal projection axes: $x$, $y$ and $z$. For each projection, we first generate the flux map from the simulation output. For each hydrodynamic cell within a cube of length $5h^{-1}$~Mpc centered on the cluster, we compute the emissivity $\epsilon_E$ for each energy $E \in [0.1,10.0]$~keV with energy bin size of $\Delta E=1$~eV, using the APEC plasma code \citep{smithetal01} with AtomDB version 2.0.2 \citep{fosteretal12}.  The emissivity $\epsilon_E = \epsilon_E(\rho,T,Z,z_{\rm obs}, v_{\rm los})$ is a function of the gas density $\rho$, temperature $T$, the gas metallicity $Z=0.3Z_\odot$, and line-of-sight velocity $v_{\rm los}$ of the cell. The observed redshift of the clusters is set to $z_{\rm obs}=0.068$, corresponding to the redshift of A1795, a nearby relaxed cluster, which is one of the likely targets for mapping out the gas velocity structure out to large radii with XARM.  We include thermal broadening in the emission lines. 

Because XARM/Resolve will be the same as ASTRO-H/SXS, each flux map is then convolved with the ASTRO-H ARF ({\tt sxt-s\_140505\_ts02um\_intall.arf}) and RMF ({\tt ah\_sxs\_5ev\_20130806.rmf}) response files from SIMX\footnote{{http://hea-www.harvard.edu/simx/}}. The energy resolution of the RMF file is $5$~eV. Photons for each location are then drawn from the convolved flux map assuming a Poisson distribution.  As we are interested in the intrinsic systematics, such as projection effects, rather than statistical uncertainties, we set the exposure time $t_{\rm exp}= 2$~Ms per pointing to ensure enough photons in the spectra. We do not model background noise because it is subdominant to the strong \FeXXV~K$\alpha$ line where the gas velocity constraints come from. 

%-------------------------------------------------%
\subsection{XARM data analyses}
%-------------------------------------------------%

In order to assess how accurately XARM can measure temperature and velocity structure in the ICM, we analyze the mock spectra and study how the physical parameters depend on various analysis conditions, fitting method, energy band, region selection, and deprojection. 
Unbinned spectra (1 spectral bin = 1~eV) are analyzed using the Cash statistic, while the data are rebinned so that each spectral bin contains at least 25 counts in the $\chi^2$  fitting. Considering the size of the Resolve field of view, the spectral regions are defined by $3\arcmin\times3\arcmin$ grids (figure~\ref{fig:CL104_x_box_region}) or annular rings with $3\arcmin$ width. The spectral fitting is performed by using XSPEC version 12.8. The spectral model consists of the BAPEC model and the Galactic absorption model \citep[the {\tt wabs} model by][]{morrison83} and is convolved by the detector and telescope responses. The gas temperature, metal abundance, redshift, velocity dispersion, and the normalization are allowed to vary while the hydrogen column density is fixed at $N_H = 2\times10^{20}~{\rm cm^{-2}}$, the value adopted when generating the mock simulations.\footnote{ The actual value of $N_H$ toward A1795 is $1.2\times10^{20}~{\rm cm^{-2}}$ according to the LAB survey of Galactic HI. This difference in $N_H $ has no consequence on our results and conclusions of the paper.}  For metal abundance, the tables in \citet{anders89} is used. The velocity dispersion is defined by the Gaussian sigma for velocity broadening. The bulk velocity along the line of sight is calculated by $c(z - z_{\rm cluster})$ where $z$ is the fitted redshift and $z_{\rm cluster} = 0.068$. Figure~\ref{fig:spec_CL104_x_2e6_cir1_3} shows a mock XARM spectrum extracted from the grid $(i,j)=(7,5)$ in figure~\ref{fig:CL104_x_box_region} of CL104 with total exposure time of 2~Msec.

The deprojection of annular spectra to recover the three-dimensional source properties is performed by the {\tt projct} model in XSPEC, assuming that the source emissivity is constant within the spherical shells whose radii correspond to the annuli used to extract the spectra. The quoted errors indicate $1\sigma$ errors. 

To quantify the accuracies of velocity measurements from the mock data, we compare them against the ``true" values measured directly from the simulations. Specifically, we compute the mass-weighted spherically averaged temperature and velocity profiles for each analyzed cluster following the procedure described in \citet{lauetal09}. Briefly, we first define the rest frame of the cluster to be the mass-weighted dark matter velocity within $r_{500}$, We then compute the gas mass-weighted mean $\langle v \rangle_{\rm mw}$ and root-mean-square velocity $\sqrt{\langle v^2 \rangle_{\rm mw}}$ in this rest-frame. We define the ``turbulent'' velocity to be the velocity dispersion $\sigma_{\rm mw} \equiv \sqrt{  \langle v \rangle_{\rm mw} ^2 - \langle v^2 \rangle_{\rm mw}}$. We compute the ``bulk'' velocity as the projected mass-weighted velocity averaged along the line-of-sight.  

%%%%%%%%%%%%%%%%%%%%%%%%%%%%%%%%%%%%%%%%%%%%%%%%%%%%%%%
\begin{figure}[t]
\begin{center}
\includegraphics[width=7cm]{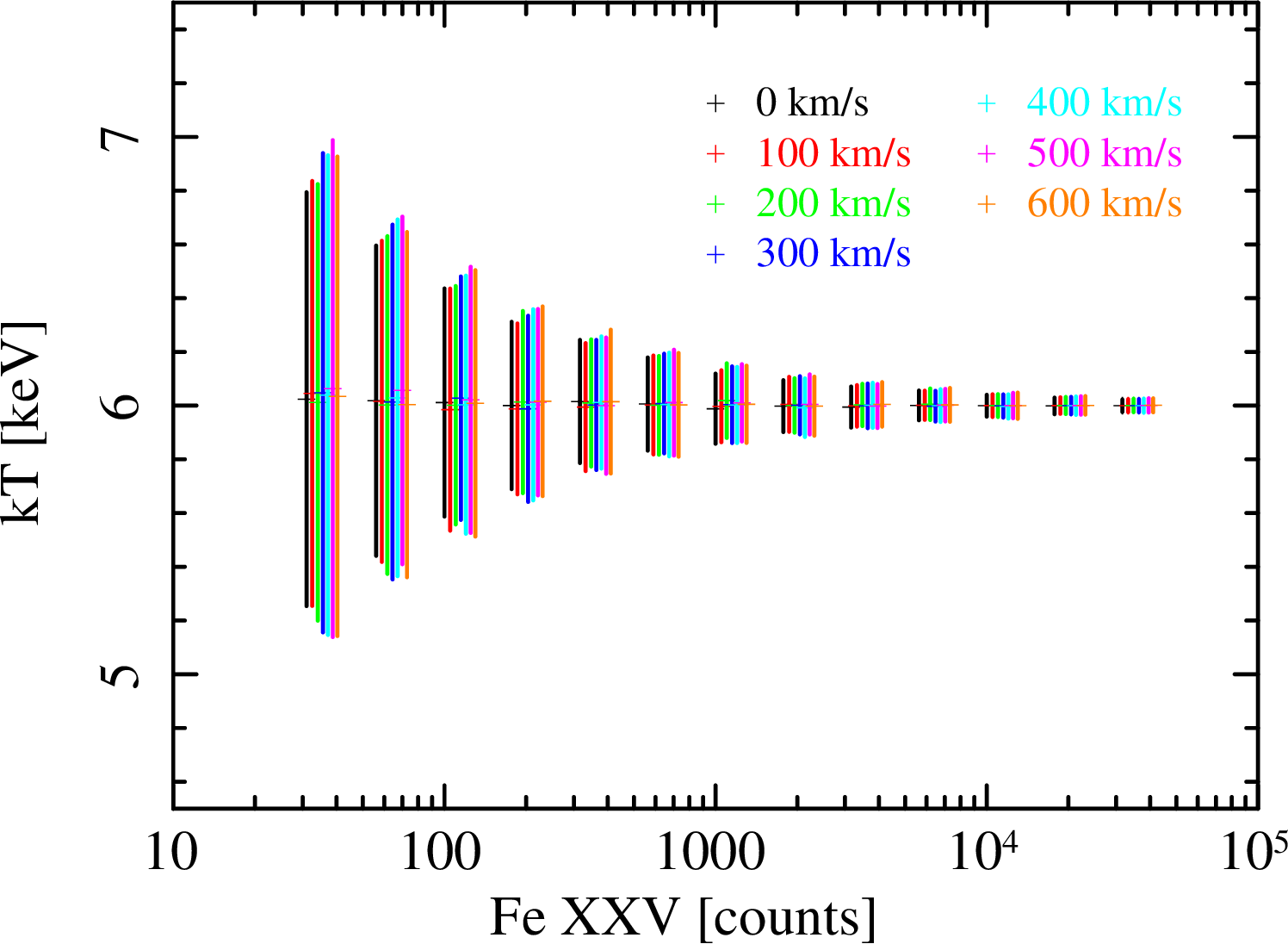}

\includegraphics[width=7cm]{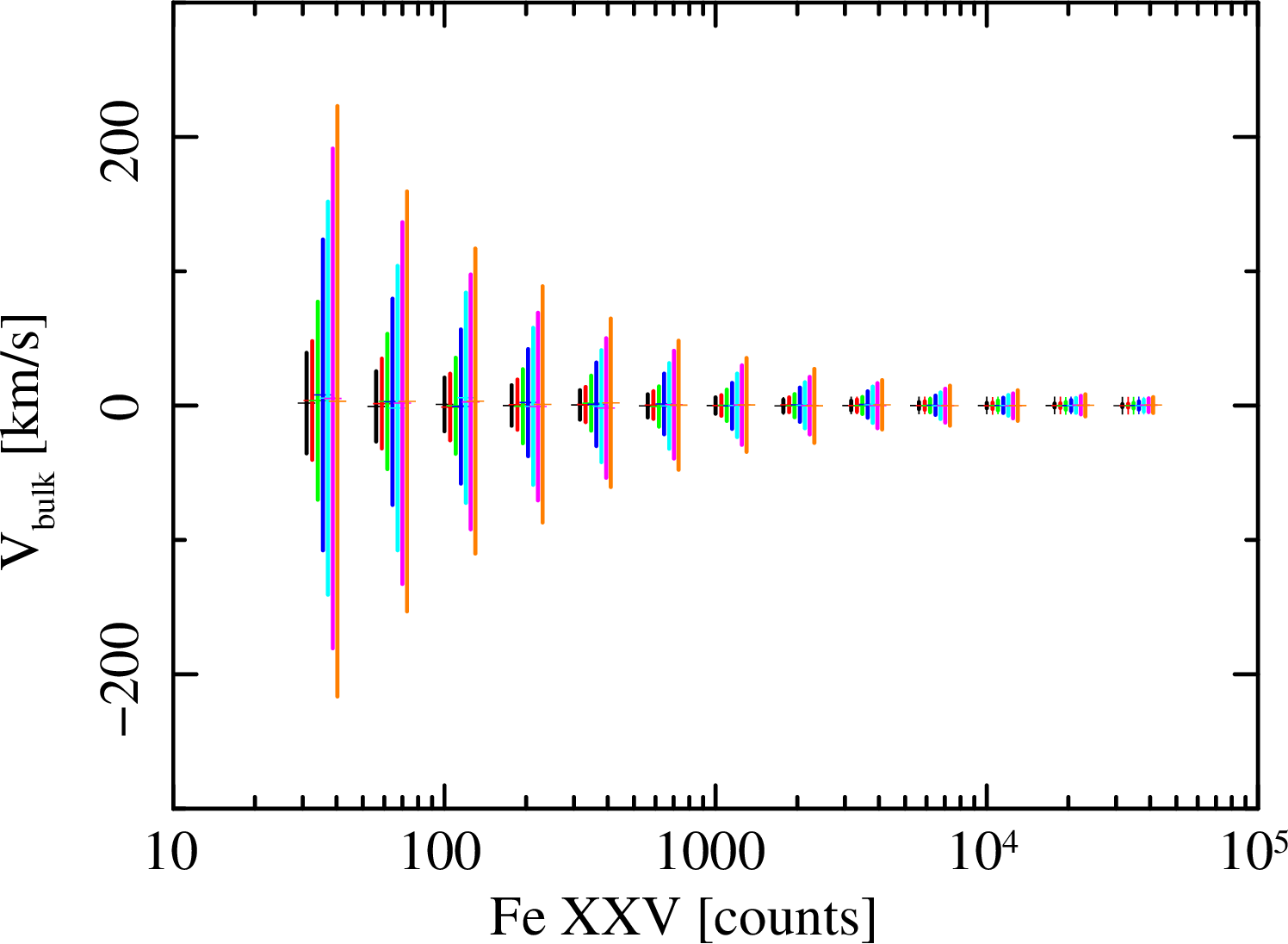}

\includegraphics[width=7cm]{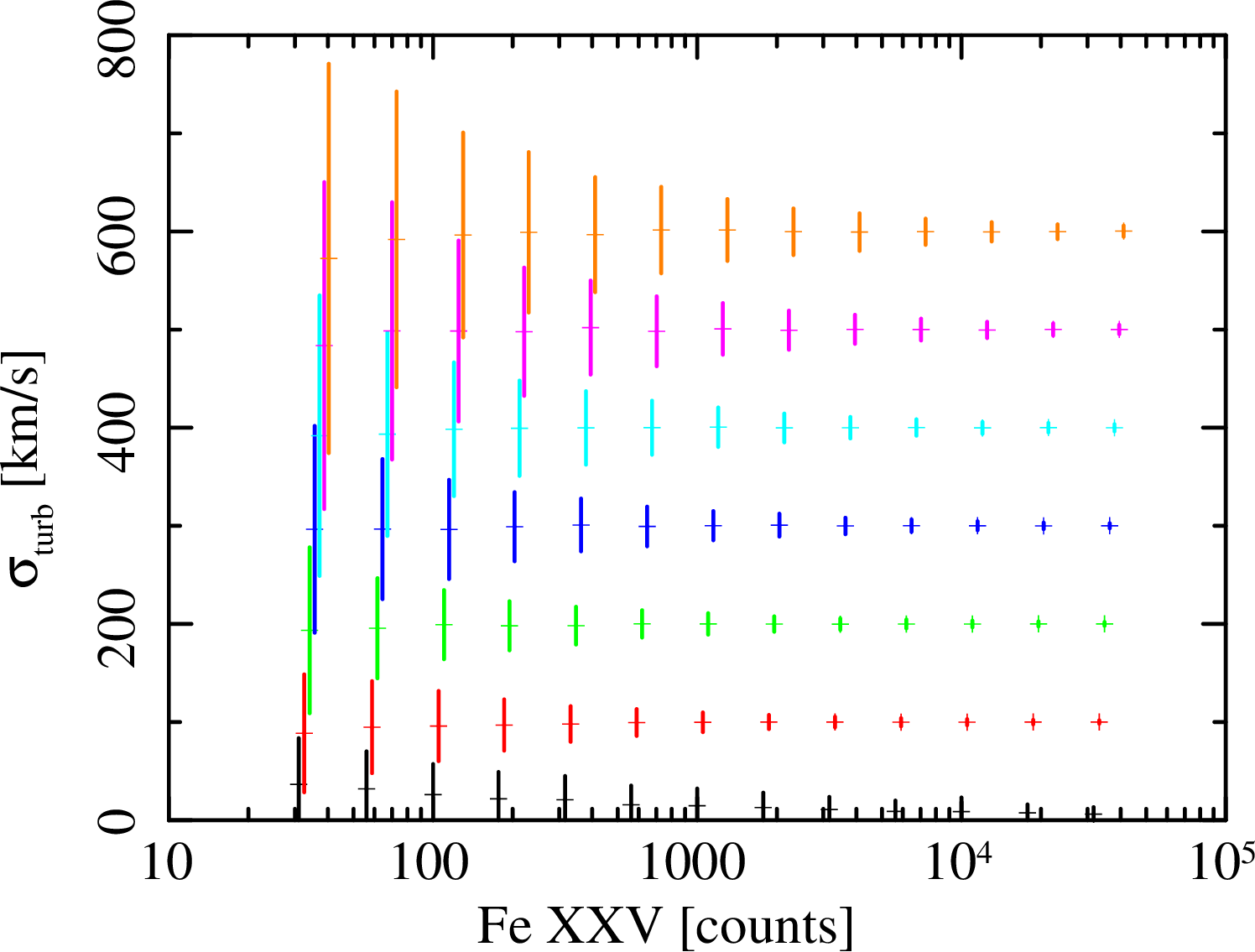}
\end{center}
\caption{Temperature (top panel), bulk velocity (middle panel), and turbulent velocity (bottom panel) as a function of counts in the \FeXXV~K$\alpha$ line complex from fitting faked spectra of the isothermal gas with $kT=6.0$~keV, $Z=0.3 Z_\odot$ in $5.0-10$~keV with the BAPEC model. Different colors indicate different input turbulent velocity $\sigma_{\rm turb}= 0$ (black), 100 (red), 200 (green), 300 (blue), 400 (cyan), 500 (magenta), and $600~\kms$ (orange).
\label{fig:counts}}
\end{figure}
%%%%%%%%%%%%%%%%%%%%%%%%%%%%%%%%%%%%%%%%%%%%%%%%%%%%%%%

%%%%%%%%%%%%%%%%%%%%%%%%%%%%%%%%%%%%%%%%%%%%%%%%%%%%%%%
\begin{figure*}[t]
\begin{center}
\includegraphics[width=5.6cm]{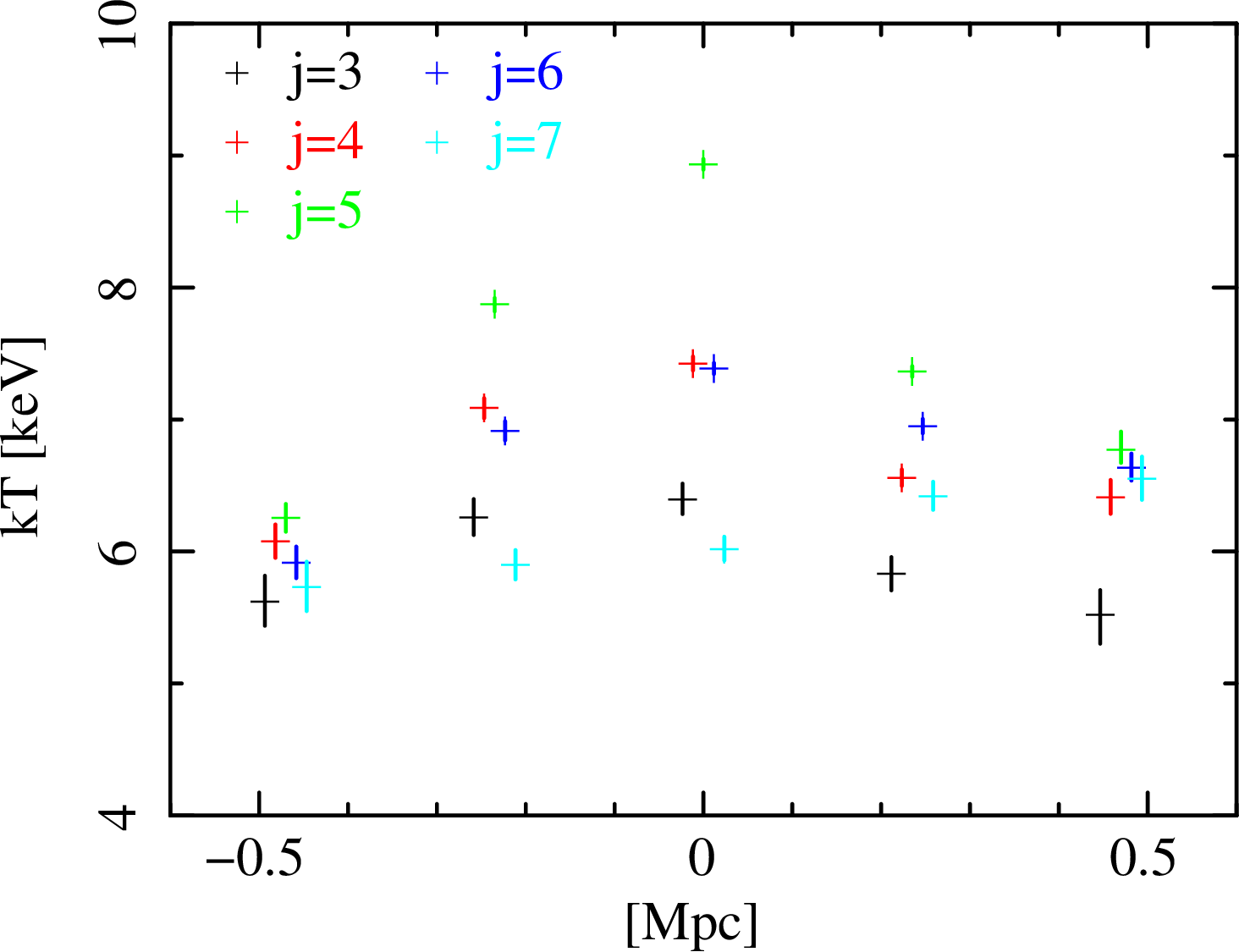}
\includegraphics[width=5.6cm]{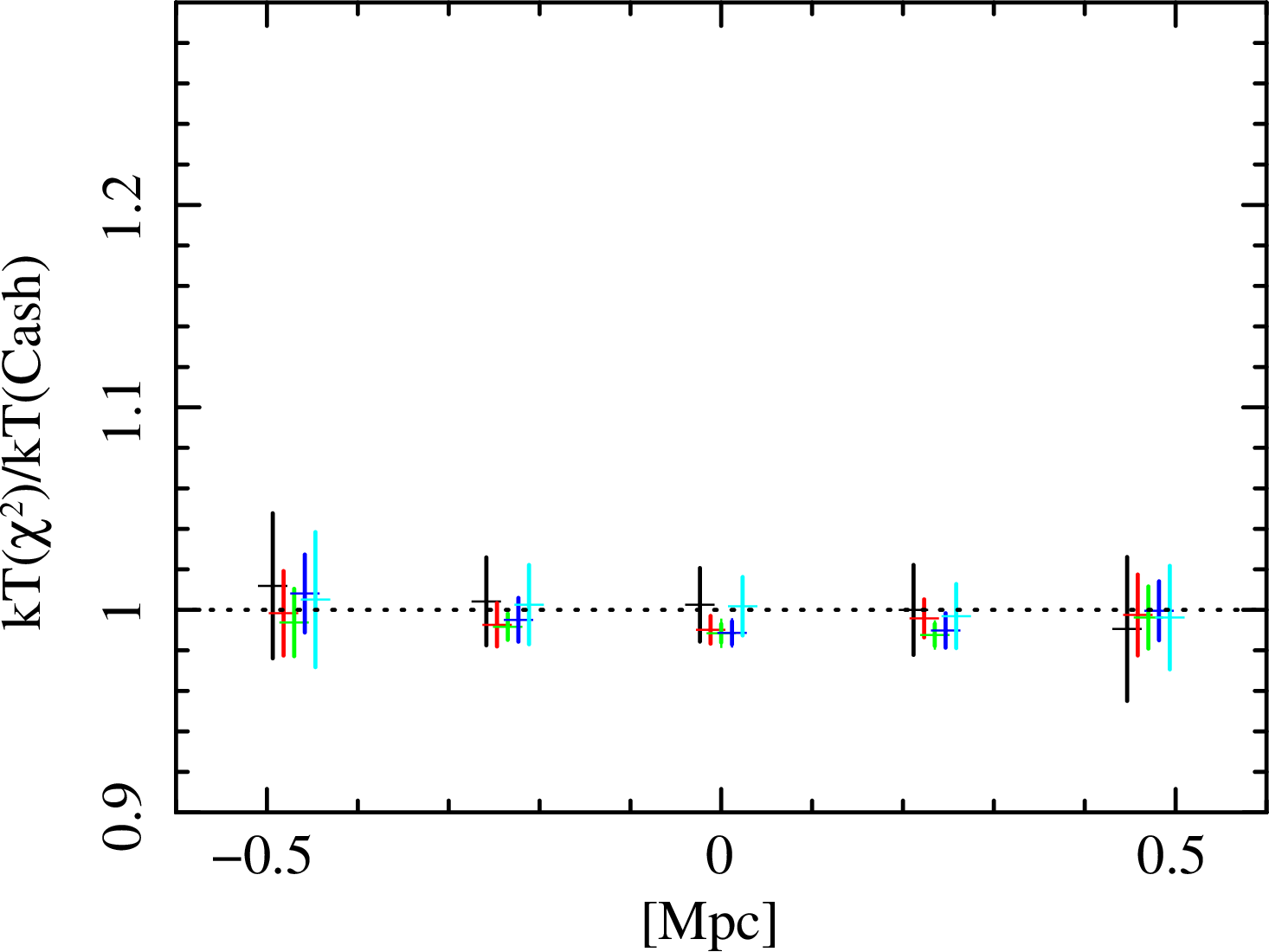}
\includegraphics[width=5.6cm]{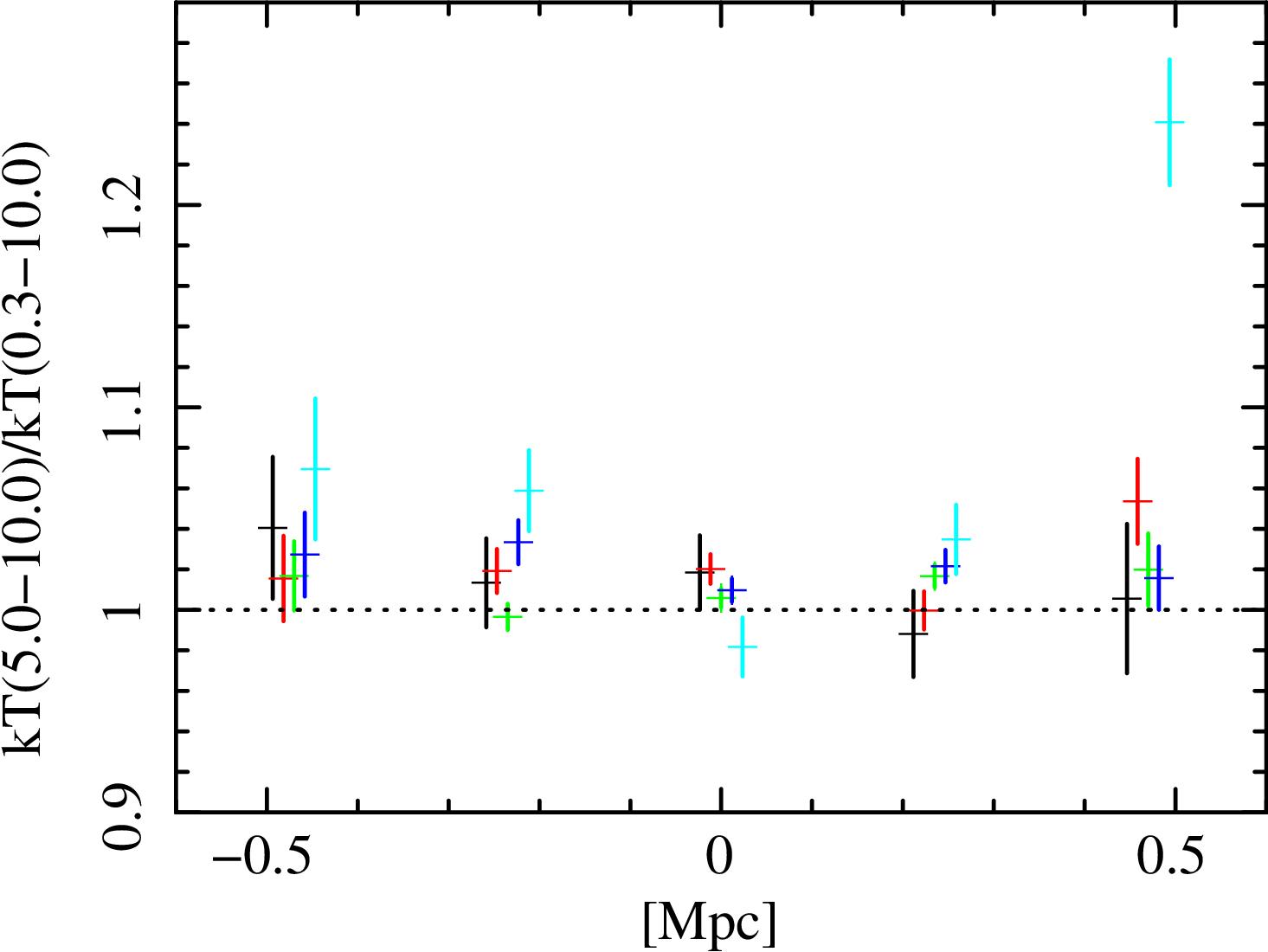}
\includegraphics[width=5.6cm]{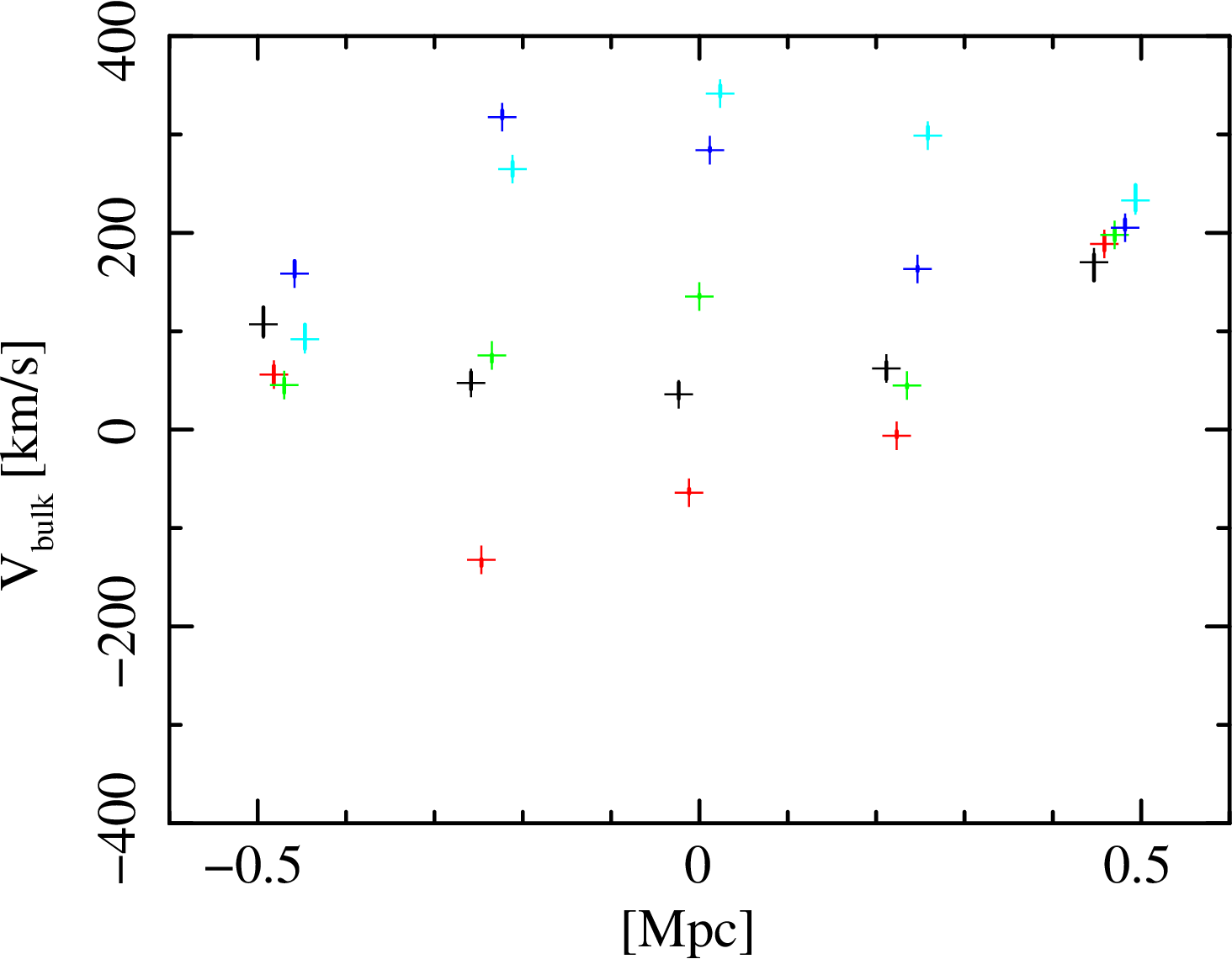}
\includegraphics[width=5.6cm]{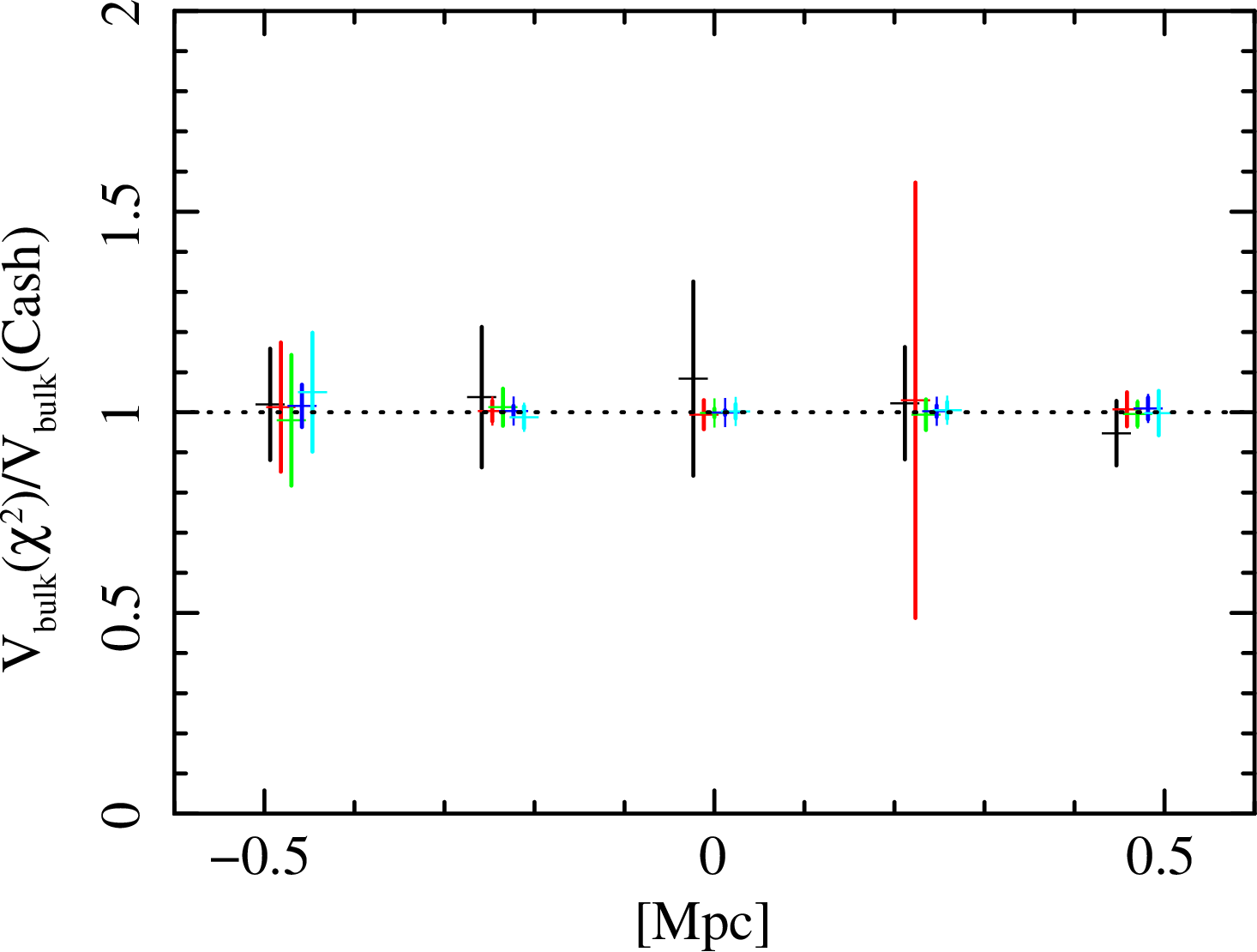}
\includegraphics[width=5.6cm]{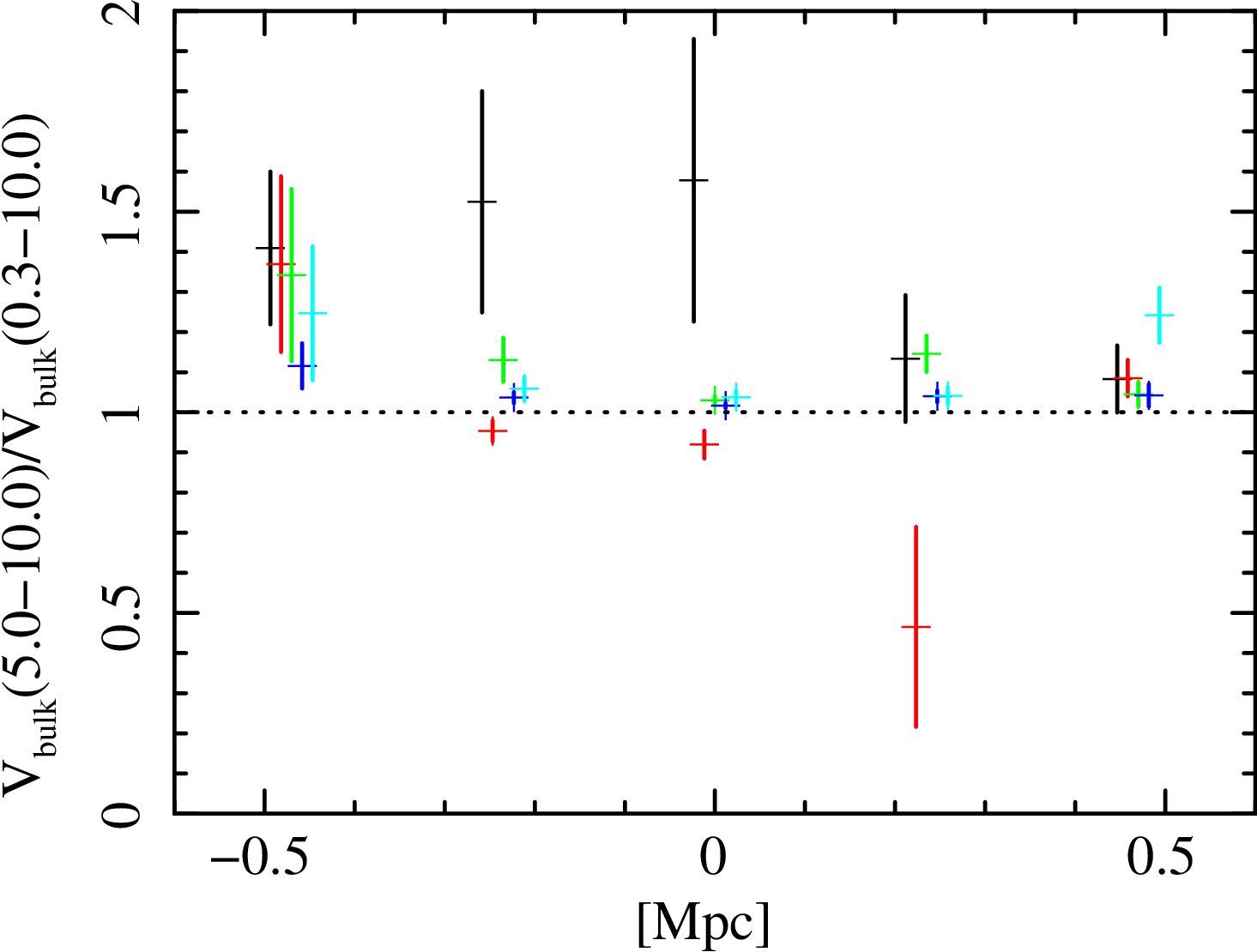}
\includegraphics[width=5.6cm]{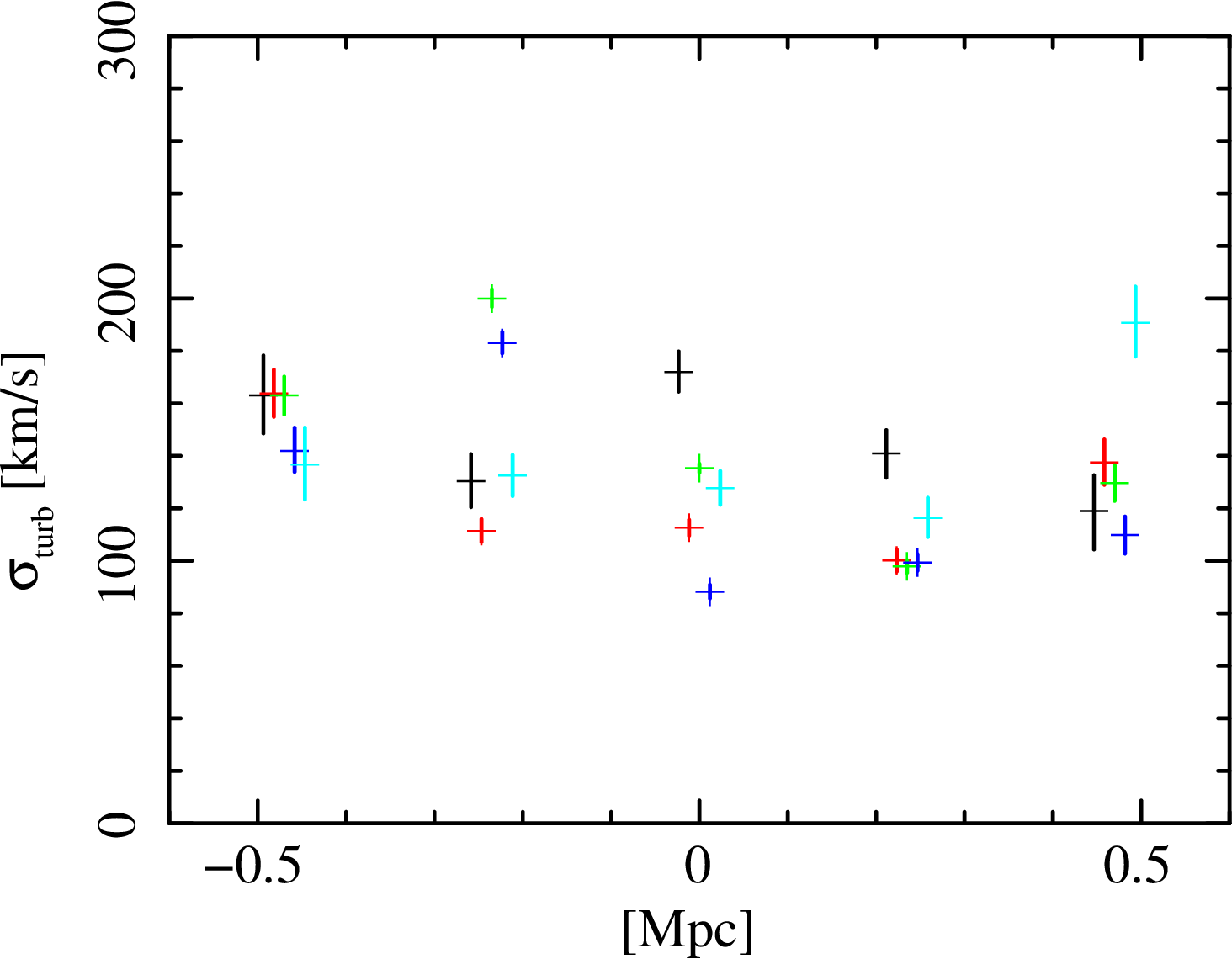}
\includegraphics[width=5.6cm]{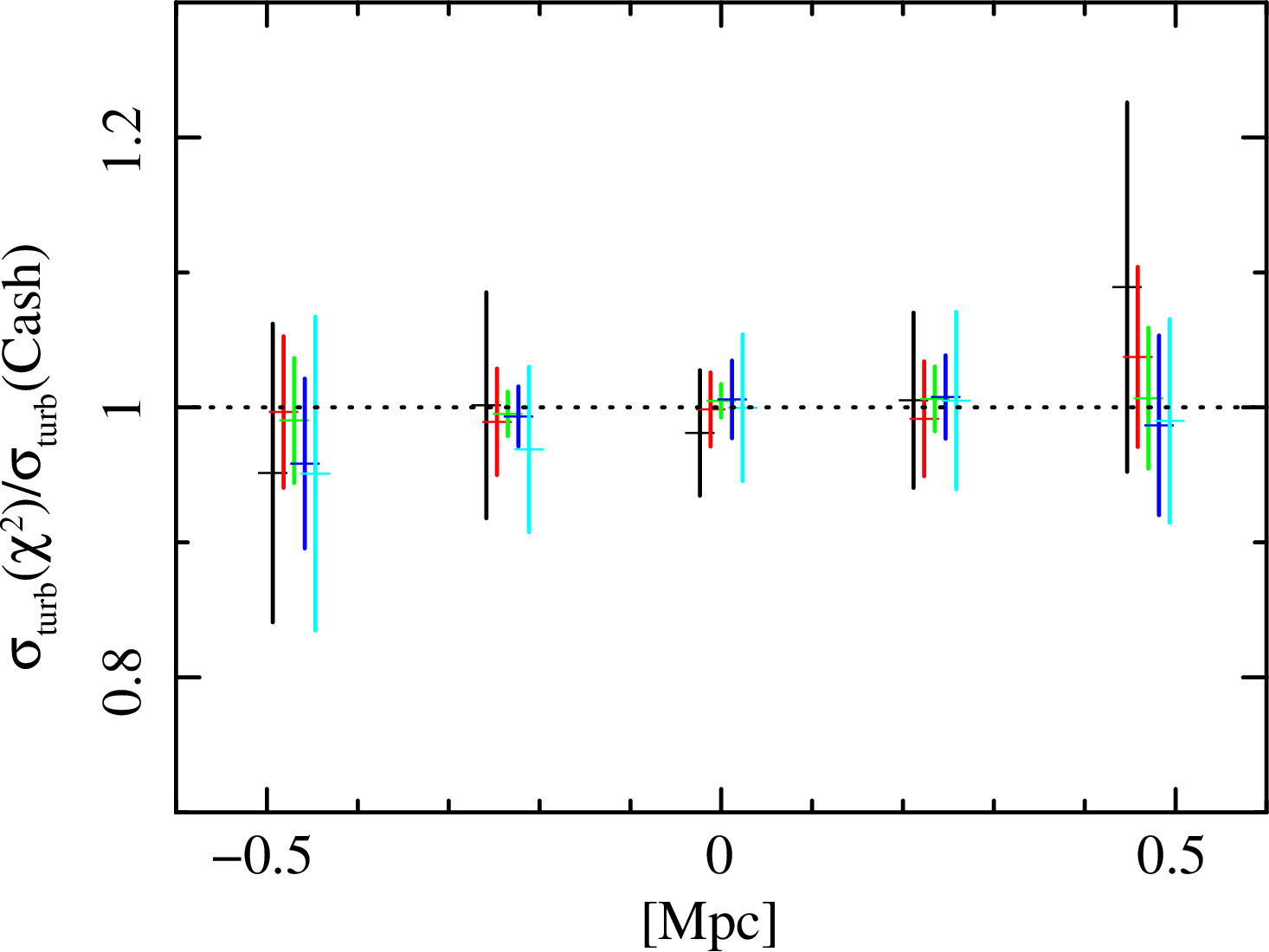}
\includegraphics[width=5.6cm]{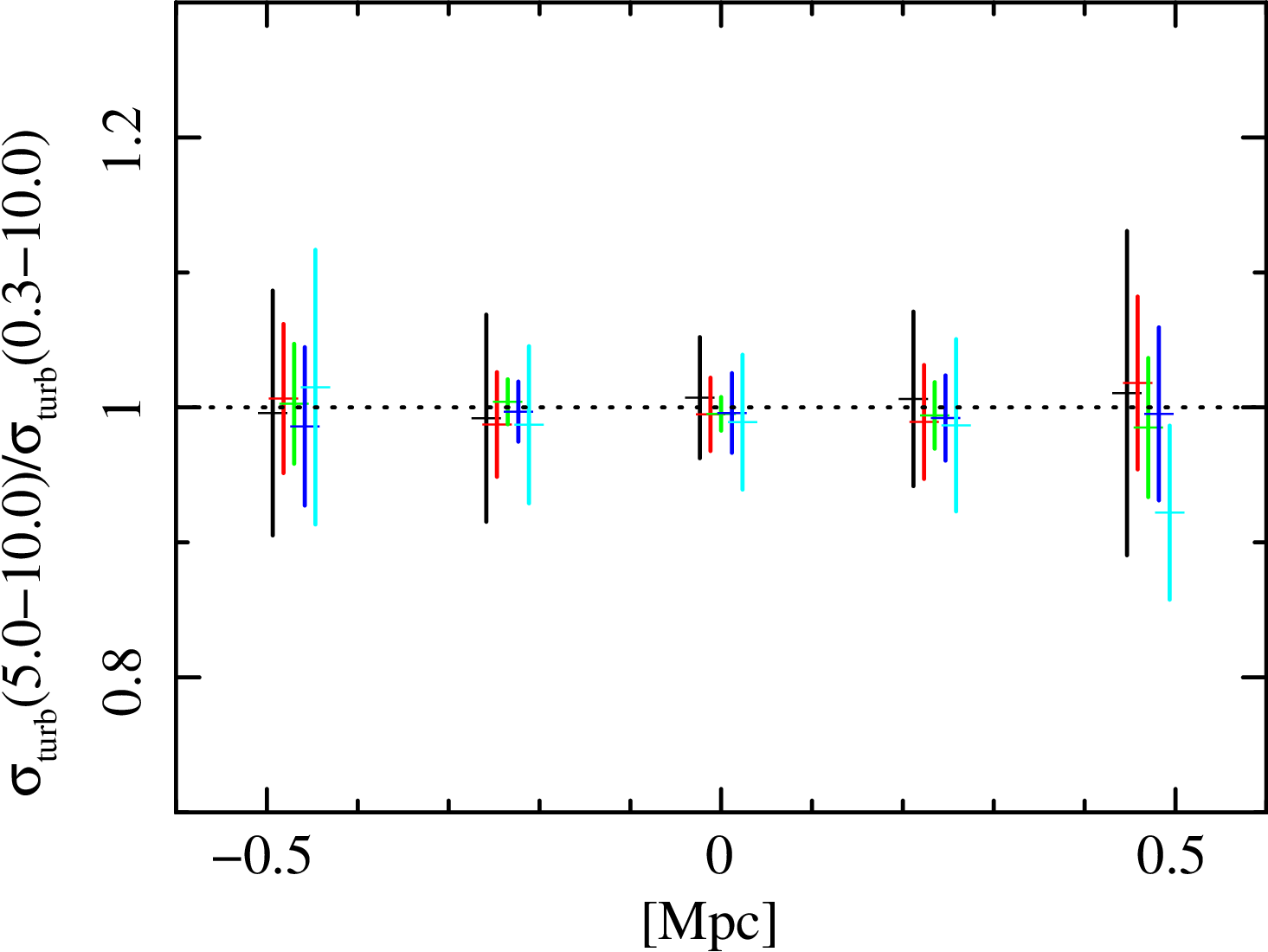}
\end{center}
\caption{Left-hand panels: results of the BAPEC model fitting to the 5.0--10.0~keV spectra using the Cash statistic for CL104. Middle panels: ratios of BAPEC model parameters from spectral fitting using the Cash and $\chi^2$ statistic in the 5.0--10~keV band for CL104. Right-hand panels: ratio of BAPEC model parameters from spectral fitting between $5.0-10$~keV and $0.3-10$~keV. The temperature (top panels), the bulk velocity (middle panels), and the turbulent velocity (bottom panels) are plotted along the $i$-axis of the map shown in figure~\ref{fig:CL104_x_box_region}. 
\label{fig:x_box_param}}
\end{figure*}
%%%%%%%%%%%%%%%%%%%%%%%%%%%%%%%%%%%%%%%%%%%%%%%%%%%%%%%

%-------------------------------------------------%
\section{Results}
\label{sec:results}
%-------------------------------------------------%

%-------------------------------------------------%
\subsection{Dependence on photon counts in the \FeXXV~K$\alpha$ line}
%-------------------------------------------------%

The primary constraints on the ICM velocity measurements with XARM come from the shifting and broadening of the bright 6.7~keV \FeXXV~K$\alpha$ line, therefore we first study the effect of photon counts in the Fe line on the parameter estimation.  We simulated Resolve spectra of isothermal gas assuming the BAPEC model with $kT=6.0$~keV, $Z=0.3 Z_\odot$, and input turbulent velocity of $\sigma_{\rm turb} = 0, 100, \dots, 600~\kms$ and zero bulk velocity.  To study the effects of photon counts, we vary the normalization so that the \FeXXV~K$\alpha$ line complex contains 30 -- 30000 counts. The simulated spectra in 5--10~keV were fitted by the BAPEC model. Figure~\ref{fig:counts} shows the best-fit parameters and their $1\sigma$ statistical errors using Cash statistic.  It shows that $\gtrsim200$ counts in the \FeXXV~K$\alpha$ line complex are required to measure the turbulent velocity accurately through line broadening for the assumed range of velocities. 
The error range in figure~\ref{fig:counts} was derived by first simulating a large number of spectra and fitted the input model to each and complied a histogram of the best-fit values from each trial and estimate the mean and the root-mean-square value. To achieve the accuracy better than 20\%, $>200$ counts are needed in the 6.7~keV \FeXXV~K$\alpha$ line complex when the turbulent velocity $\sigma_{\rm turb} < 200~\kms$. We recommend $>200$ counts in 6.7~keV \FeXXV~K$\alpha$ line complex for robust measurements of $\sigma_{\rm turb}$.
Note that the requirement for calibration accuracy of SXS energy-scale is 2~eV with a goal of 1~eV \citep{mitsudaetal14}, which corresponds to the line shift of $90~\kms$ (or $45~\kms$) at 6.7~keV. Thus, in very bright regions like cluster cores, the measurement is limited by the instrumental uncertainty rather than the statistical error.

%-------------------------------------------------%
\subsection{Fitting statistics: Cash vs.\ $\chi^2$}
%-------------------------------------------------%

Two statistics, Cash \citep{cash79} and $\chi^2$, are commonly used in X-ray spectral fitting. We compare how these two statistics differ in the parameter estimation for the relaxed cluster CL104. Because velocity measurement requires sufficient photon counts ($>200$ counts) in the \FeXXV~K$\alpha$ line, the spectral regions are limited to the central $5\times5$ grids (approximately $r<r_{2500}$). The left-hand panels of figure~\ref{fig:x_box_param} shows results of the BAPEC model fitting to the 5.0--10.0~keV spectra using the Cash statistic and the middle panels shows comparison of resultant parameters between Cash and $\chi^2$ statistics. Each spectrum contains at least 200 counts in the \FeXXV~K$\alpha$ line. The differences in the fitted temperature, bulk and turbulent velocities between the two different methods are typically less than $20\%$, which is reasonable as the Cash statistic approaches the $\chi^2$ statistic in the limit of large photon counts. Since two methods produce consistent results, the unbinned spectra are analyzed by utilizing the Cash statistic in the following analysis. Cash statistic is preferred for low photon counts, the spectrum does not need to be rebinned as in the case of $\chi^2$ statistic, thus preserving the spectral energy resolution crucial for measuring gas velocities from line broadening. 

%-------------------------------------------------%
\subsection{Spectral energy bands and multiphase ICM}
%-------------------------------------------------%

%%%%%%%%%%%%%%%%%%%%%%%%%%%%%%%%%%%%%%%%%%%%%%%%%%%%%%%
\begin{figure}
\begin{center}
\includegraphics[width=7cm]{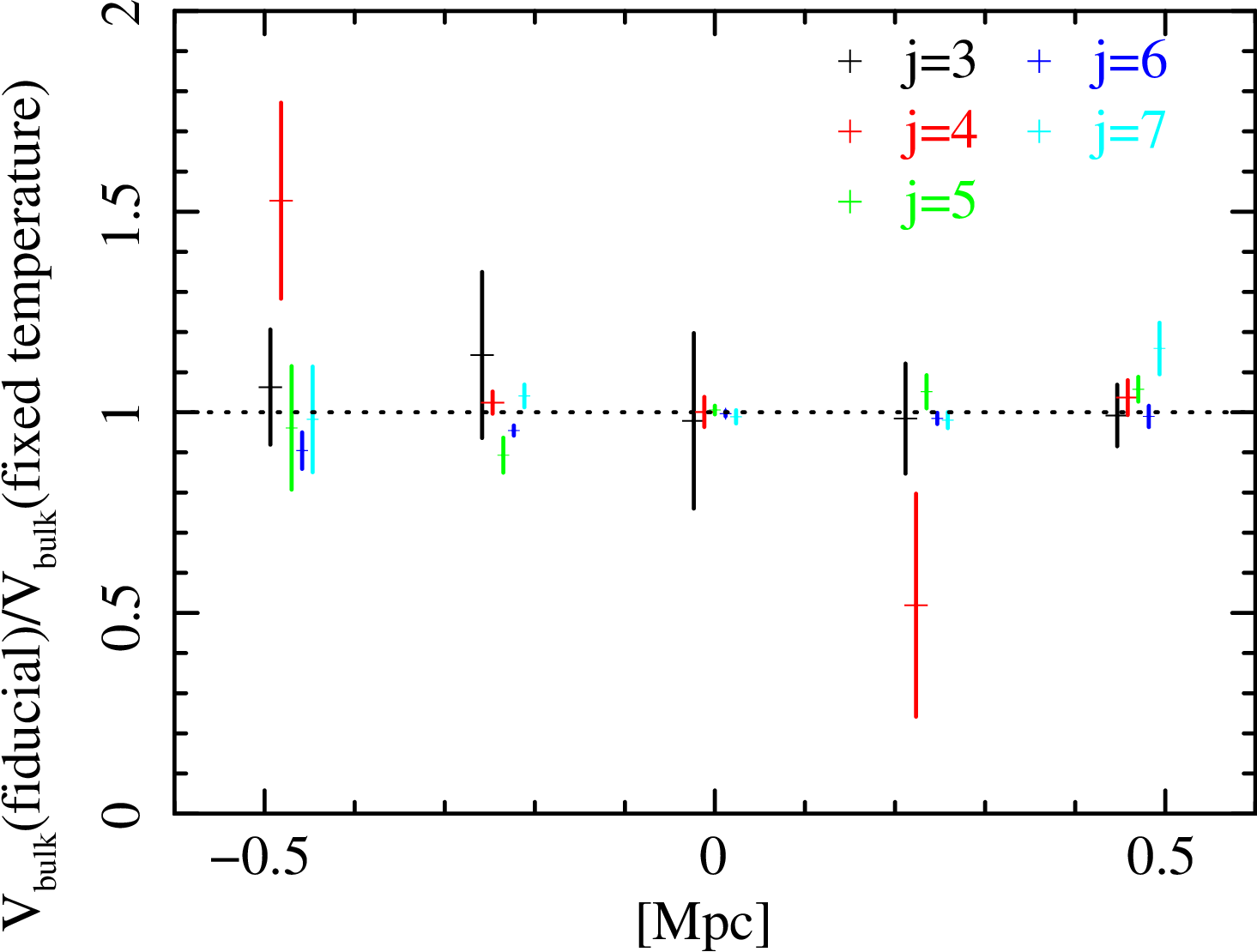}

\includegraphics[width=7cm]{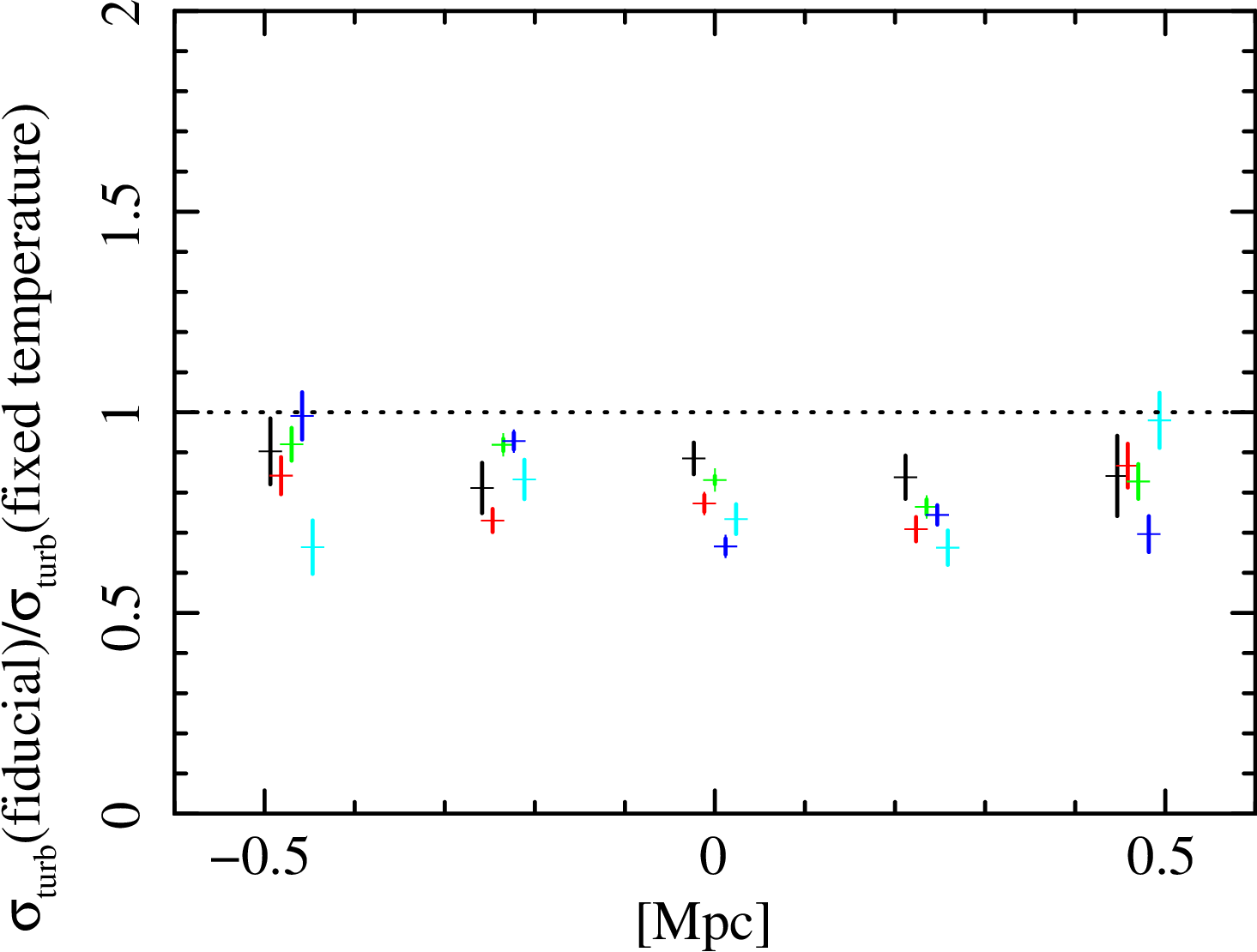}
\end{center}
\caption{Comparison of bulk (top) and turbulent (bottom) velocities derived in the fiducial case with the true gas temperature and the fixed temperature case with $T_X = 7.7$~keV, for CL104. Different color points correspond to different $j$ location in the map shown in figure~\ref{fig:CL104_x_box_region}. \label{fig:ftemp}}
\end{figure}
%%%%%%%%%%%%%%%%%%%%%%%%%%%%%%%%%%%%%%%%%%%%%%%%%%%%%%%

The temperature of the ICM may affect the accuracies of turbulent velocity measurements from line broadening. The line broadening originates from both thermal and turbulent motions and given by a combination of $\Delta E_{\rm th} = E_0 (kT/m)^{1/2}c^{-1}$ and $\Delta E_{\rm turb} = E_0 \sigma_{\rm turb}c^{-1}$, where $E_0$ is the rest-frame energy of the line emission, and $m$ is the ion mass. Since the temperature $kT$ is determined primarily by the shape of continuum spectrum of the hot ICM, it is ideal to use a wider energy range for accurate parameter estimation. If a certain amount of cool gas exists along the line of sight, the soft emission may  affect the temperature determination even though the emissivity of \FeXXV~K$\alpha$ line at low temperature is low. We thus compare analysis in two different energy ranges, 5.0--10~keV and 0.3--10~keV, where the latter includes more emission from cool gas. 

As seen from the right-hand panels in figure~\ref{fig:x_box_param}, for the relaxed cluster CL104, the fitted parameters do not depend on the choice of the energy range for most regions except for the grid $(i, j )= (7,7)$, where the temperature derived from 5.0--10~keV is $\sim 25\%$ higher than that derived from 0.3--10~keV. This is due to the presence of an infalling cooler substructure which is visible in the north-west of the cluster image, shown in figure~\ref{fig:CL104_x_box_region}. 

The recovered turbulent velocity does not show clear dependence on the energy range. This is expected as most of the constraints on the turbulent velocity come from the broadening of the \FeXXV~K$\alpha$ line present in both energy ranges.  Since the velocity constraints from line broadening is inversely proportional to the mass of the ion species, velocity measurements based on the Doppler broadening of the \FeXXV~K$\alpha$ ion is least affected by thermal broadening compared to emission lines of species of lighter elements 
\citep[e.g.,][]{inogamovsunyaev03}. On the other hand, the soft band (0.3--5.0~keV) contains emission lines from lighter elements which are more susceptible to thermal broadening, and leads to biases in turbulent velocity estimates.  

We further examine the effects of multiphase ICM by fixing the gas temperature to $7.7$~keV, while keeping the gas density, abundance, exposure time and fitting method unchanged. The velocities in this case should be unbiased from the effects of the multi-temperature ICM. We compare the velocity estimates between the fixed temperature case and the fiducial, multi-temperature case in figure~\ref{fig:ftemp}. While the bulk velocity measurements of the two cases are consistent with each other within statistical error bars, the turbulent velocity is underestimated by $\sim 30\%$ in the fiducial case compared to the fixed temperature case. This is expected because the turbulent velocity (determined from broadening of the \FeXXV~K$\alpha$ line) is more affected by higher temperature gas through thermal broadening, while the bulk velocity (determined from line shift) is not.

%%%%%%%%%%%%%%%%%%%%%%%%%%%%%%%%%%%%%%%%%%%%%%%%%%%%%%%
\begin{figure}
\begin{center}
\includegraphics[width=4.0cm]{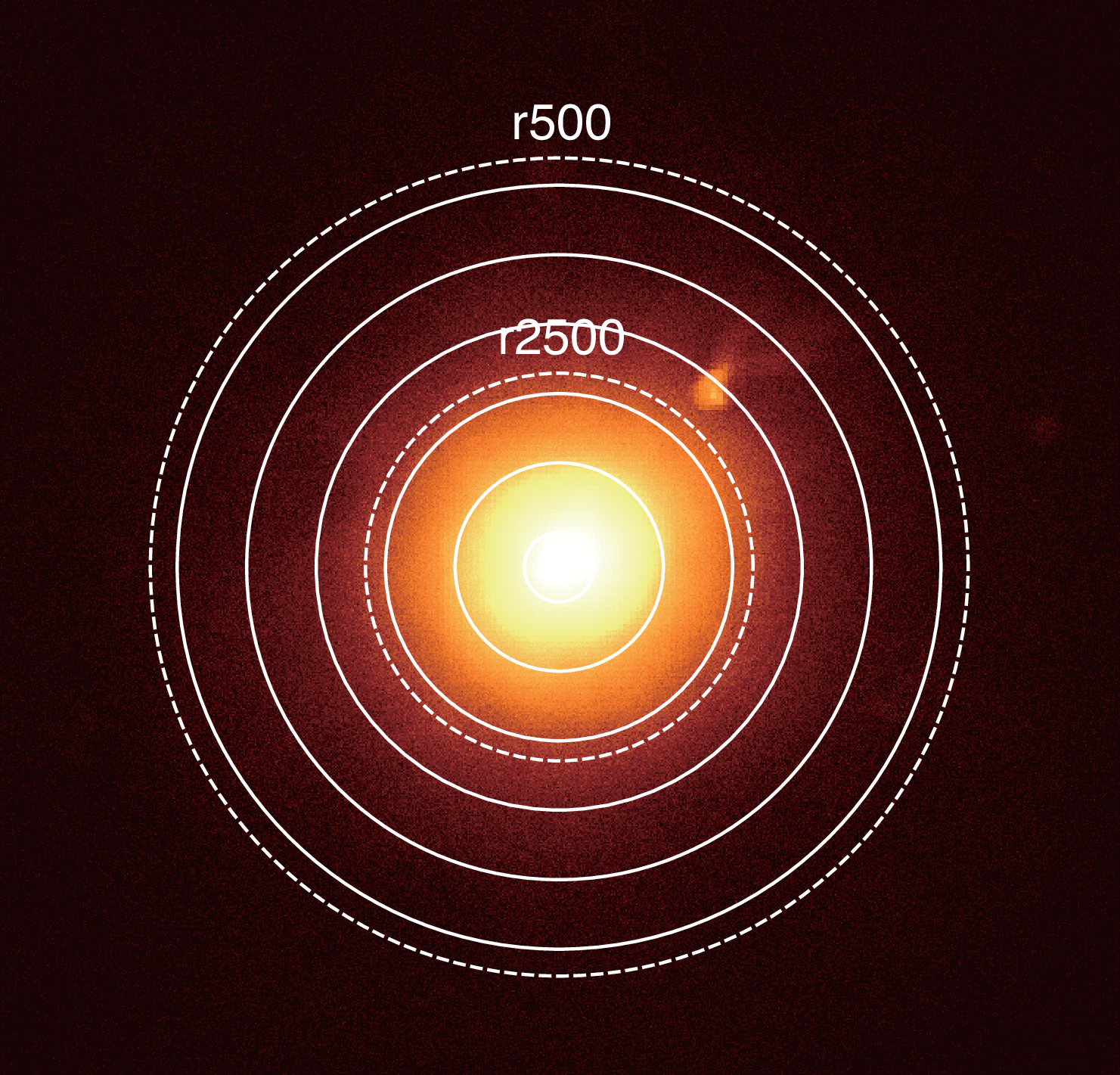}
\includegraphics[width=4.0cm]{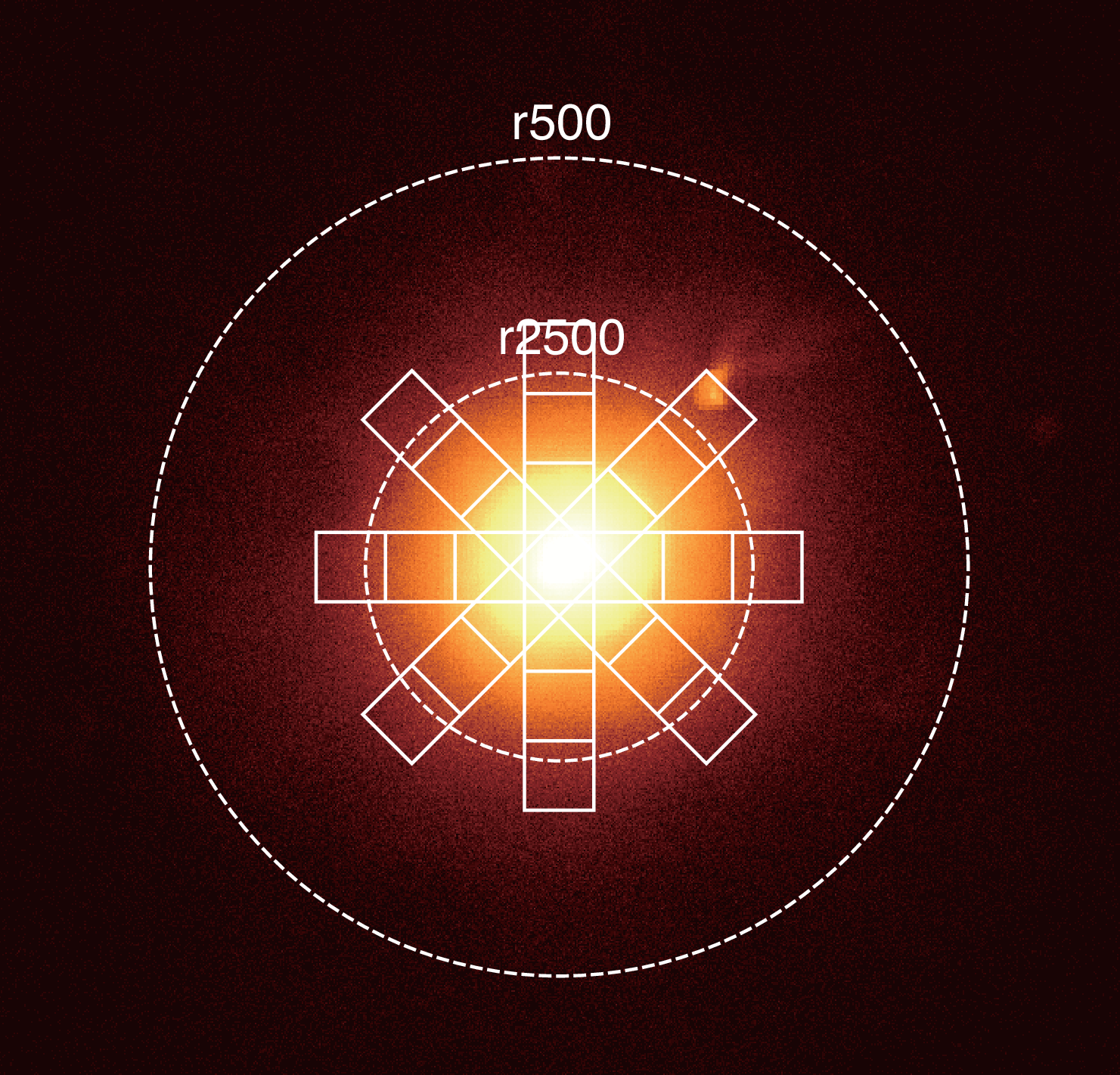}

\includegraphics[width=4.0cm]{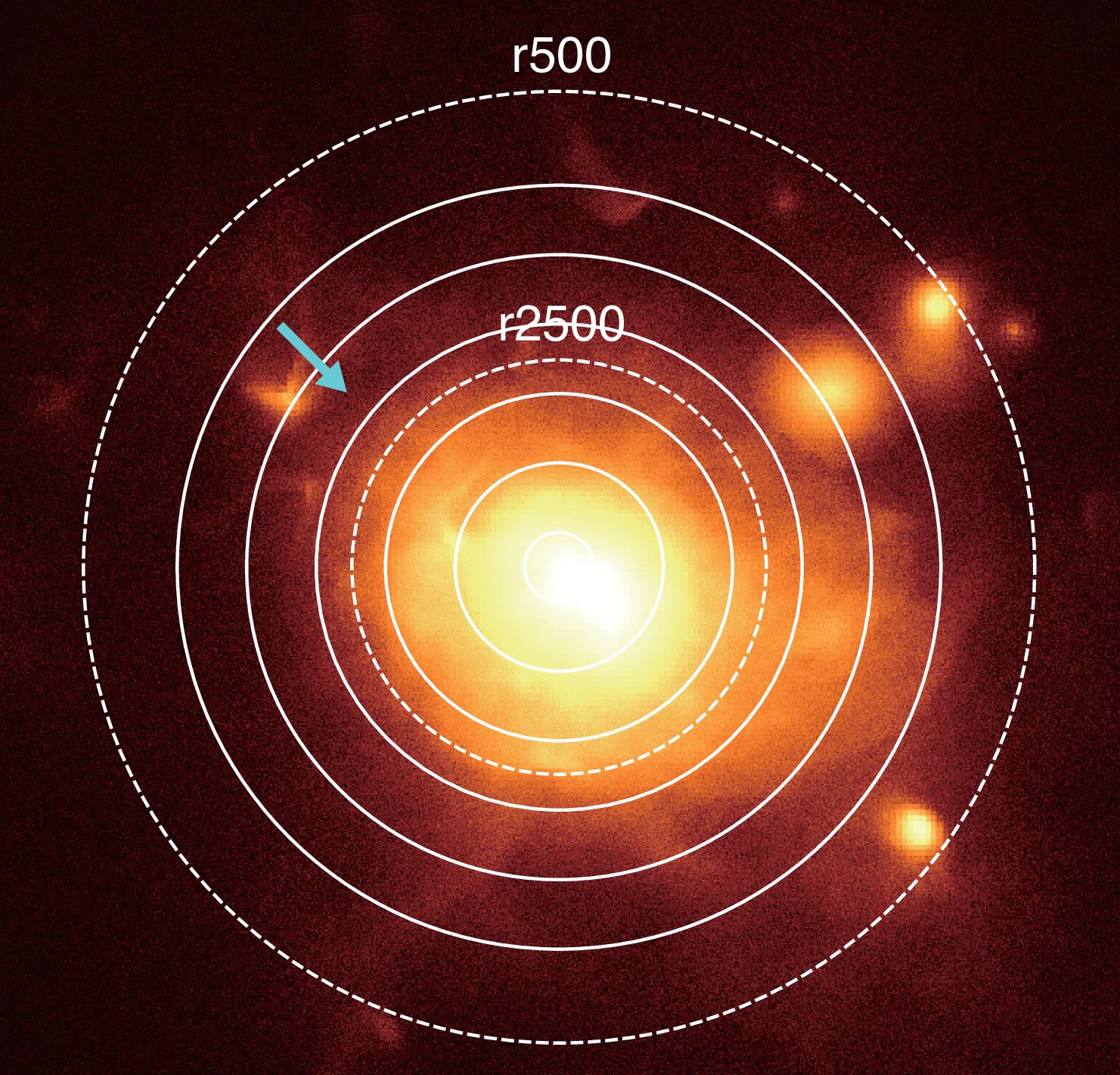}
\includegraphics[width=4.0cm]{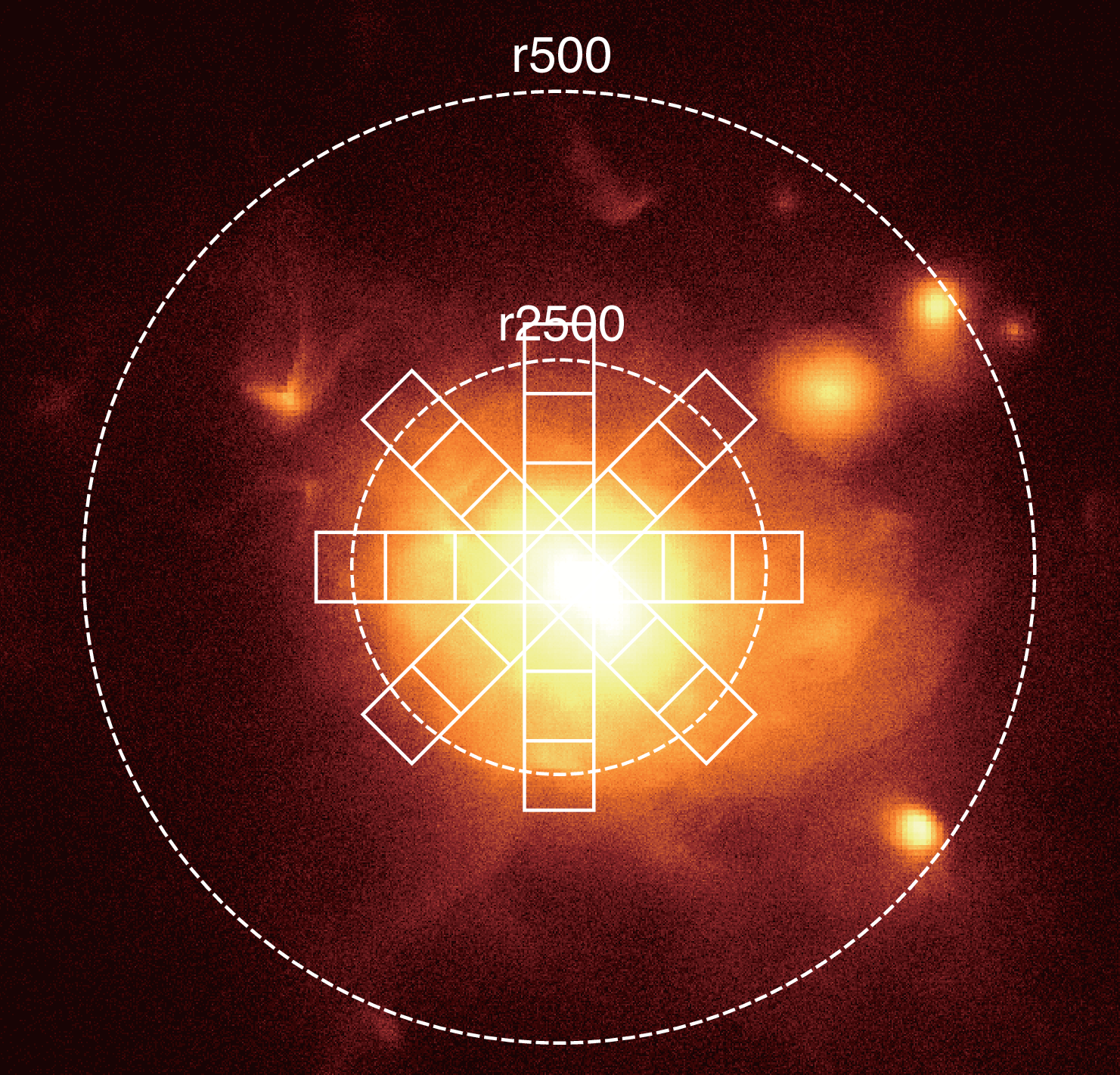}
\end{center}
\caption{Top panels: spectral regions for the deprojection analysis for CL104: six annuli (left) and four boxes in eight azimuthal directions (right).  Bottom panels: the same regions are shown for CL101.  The maps show the $x$-projections for the two clusters. The location of merger shock is indicated by the arrow (see subsection~\ref{subsubsec:dynamicalstate}).  \label{fig:region}
}
\end{figure}
%%%%%%%%%%%%%%%%%%%%%%%%%%%%%%%%%%%%%%%%%%%%%%%%%%%%%%%

%%%%%%%%%%%%%%%%%%%%%%%%%%%%%%%%%%%%%%%%%%%%%%%%%%%%%%%
\begin{figure*}
\begin{center}
\includegraphics[width=5.5cm]{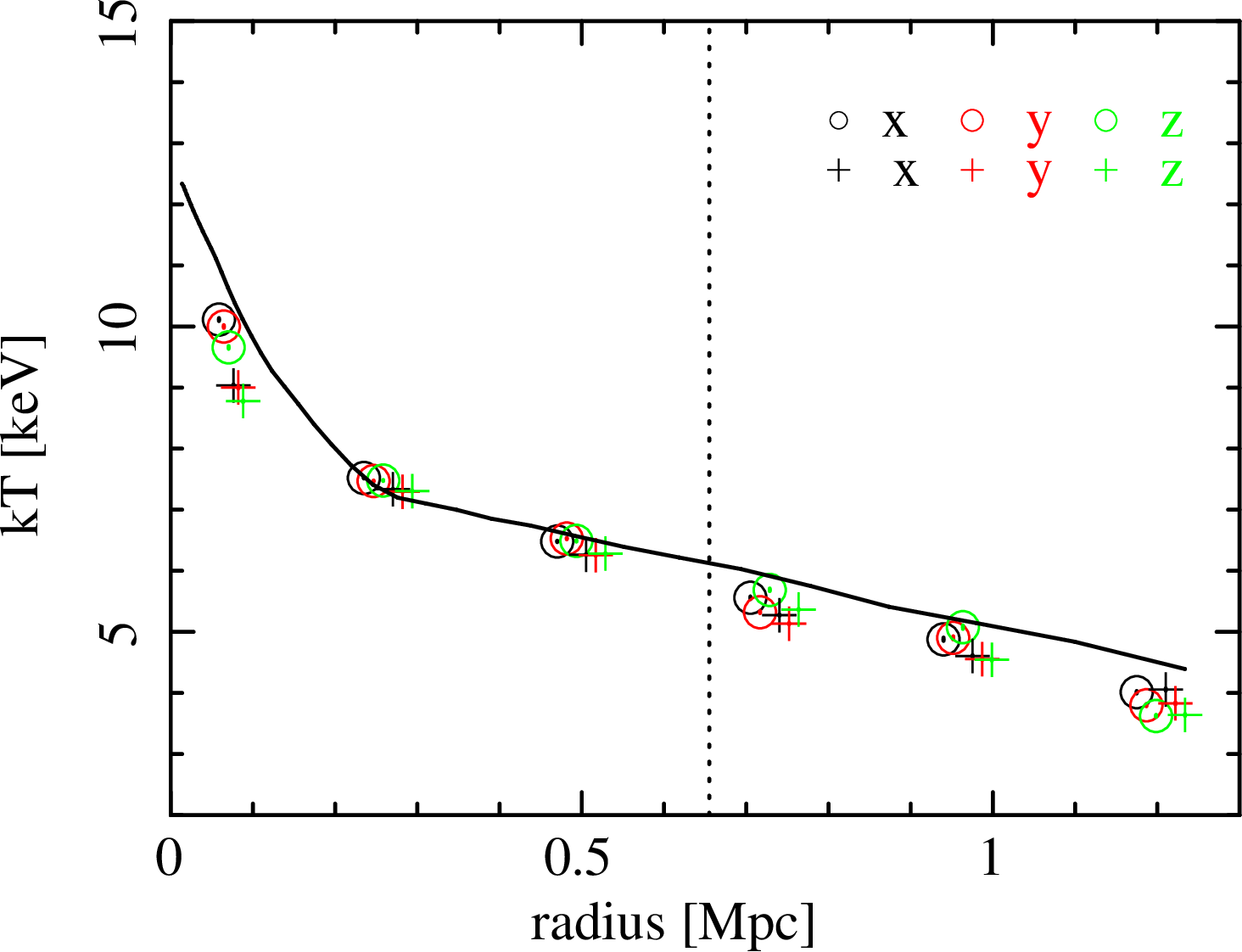}
\includegraphics[width=5.5cm]{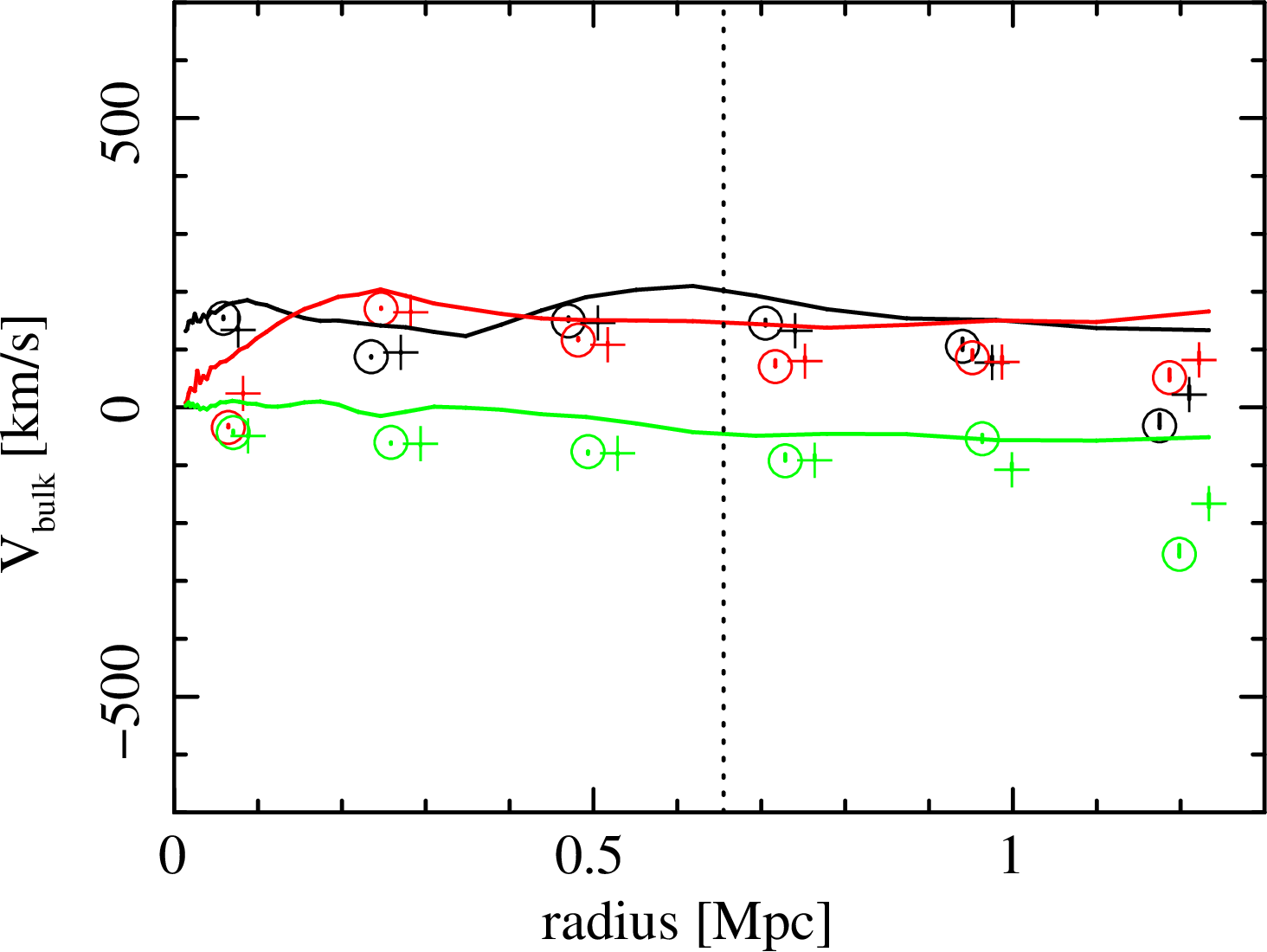}
\includegraphics[width=5.5cm]{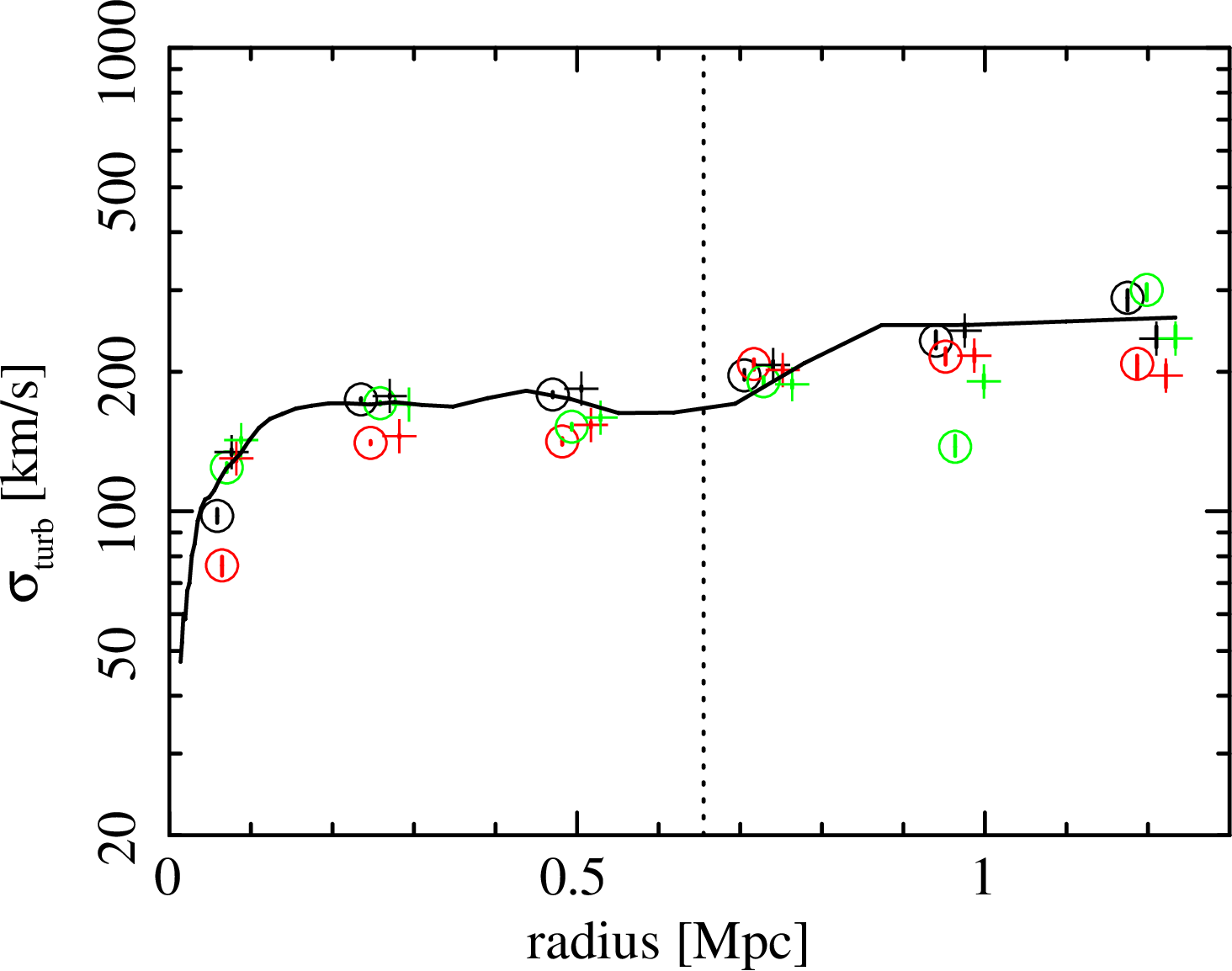}
\includegraphics[width=5.5cm]{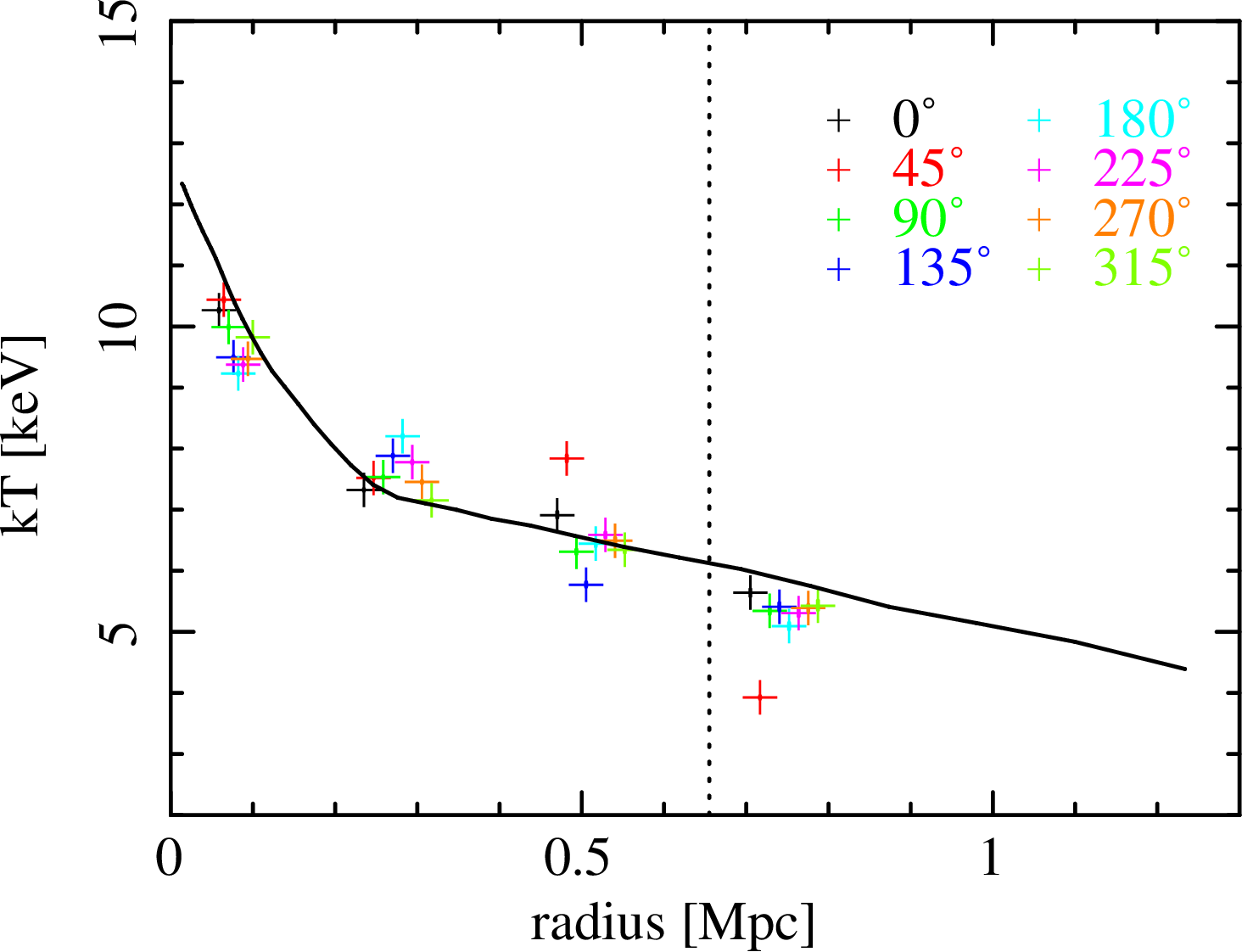}
\includegraphics[width=5.5cm]{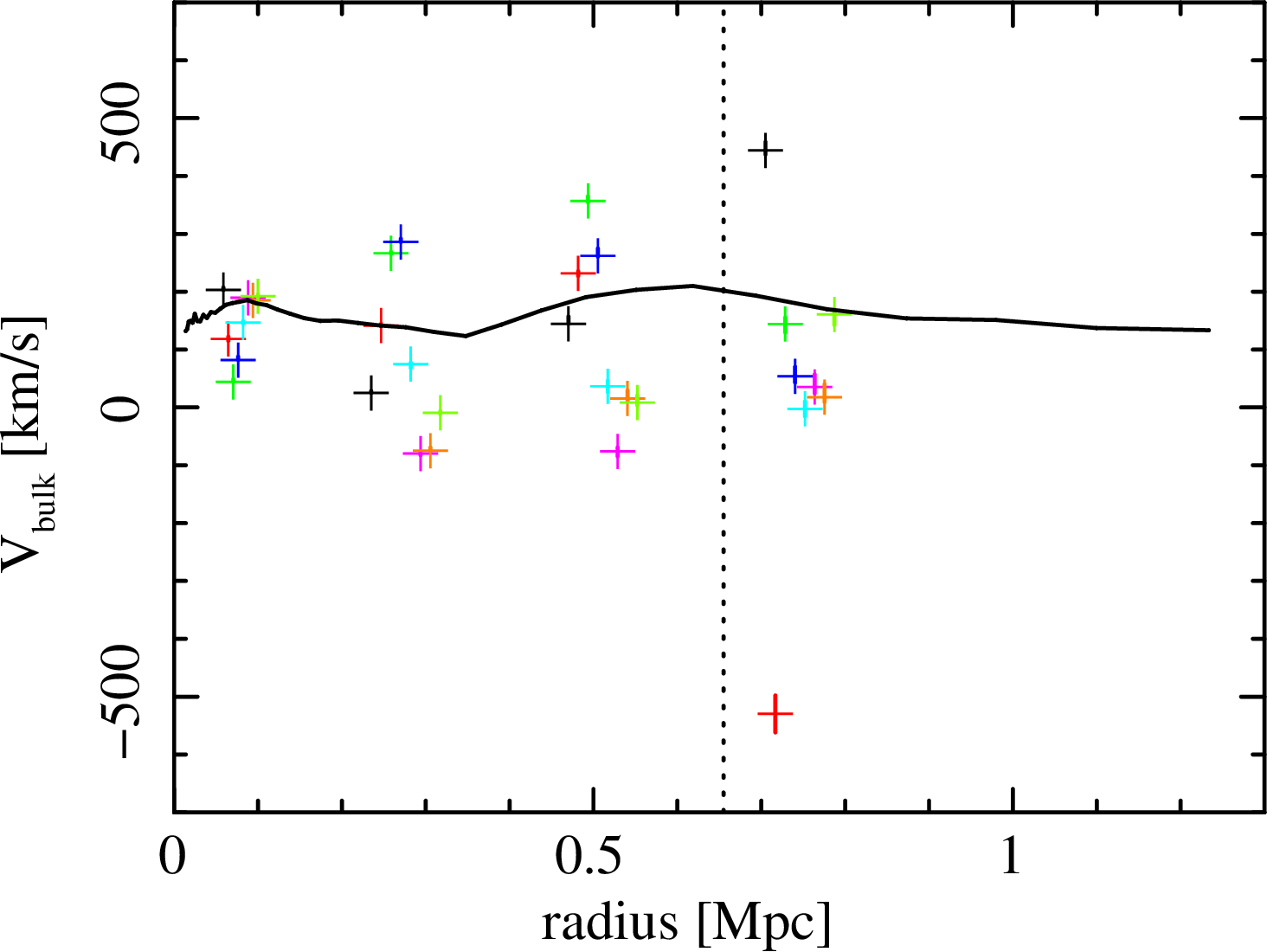}
\includegraphics[width=5.5cm]{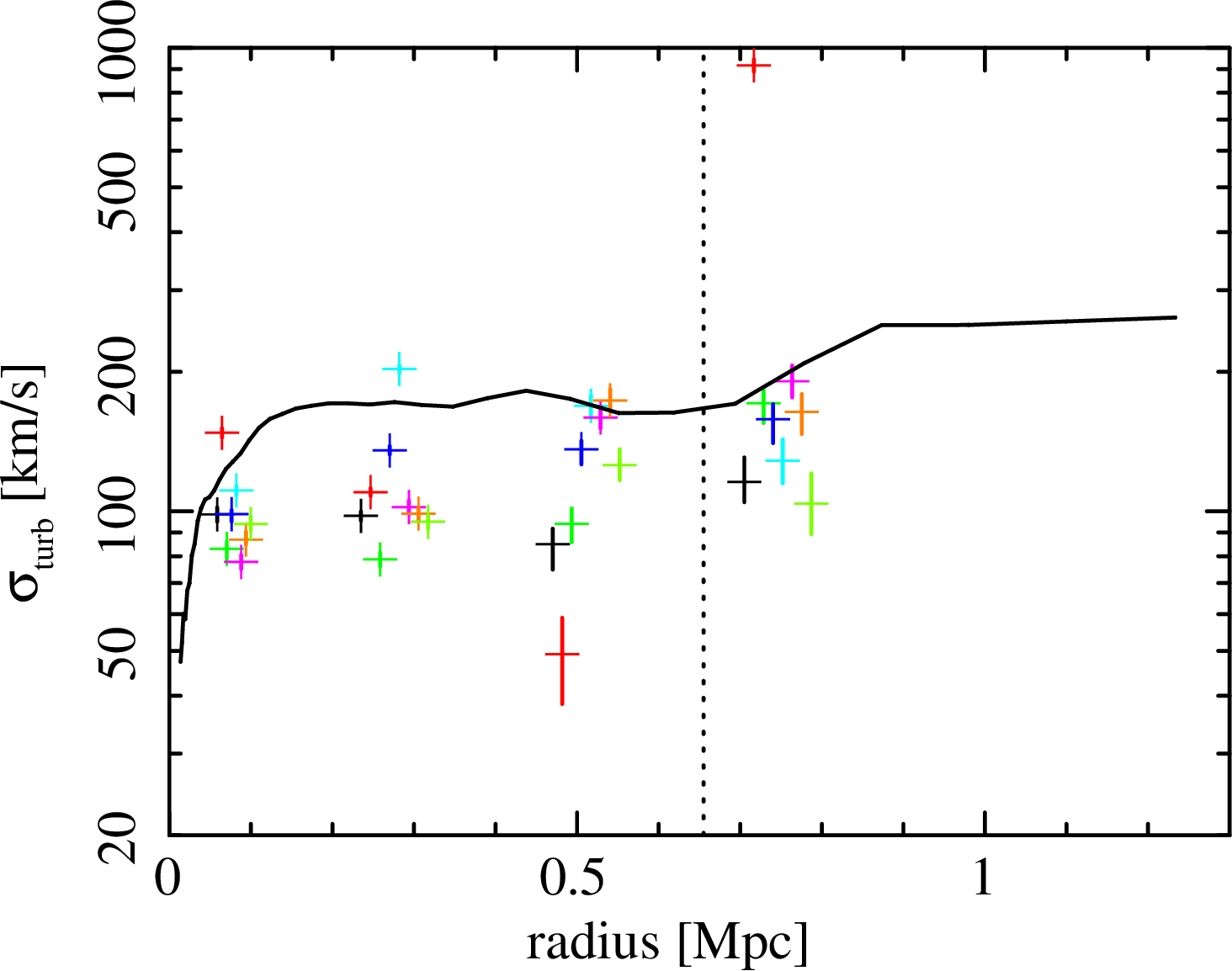}
\end{center}
\caption{Top panels: Deprojected temperature (left-hand panels), bulk velocity (middle panels), and turbulent velocity (right-hand panels) of the relaxed cluster CL104 viewed along three orthogonal projections: x (black), y (red), z (green) projections (open circles). The projected quantities are shown with the crosses for comparison. 
Bottom panels: Deprojected temperature, bulk velocity, and turbulent velocity for the eight azimuthal directions (crosses) for the x projection.  The Cash statistic was used to fit the $0.3-10$~keV spectra. 
The solid lines indicate mass-weighted, spherically averaged values for temperature and velocity dispersion, and the mass-weighted projected velocity along the line-of-sight,  computed directly from simulation. The dotted line indicates $r_{2500}$ of the cluster. 
\label{fig:deproj_circle}
}
\end{figure*}
%%%%%%%%%%%%%%%%%%%%%%%%%%%%%%%%%%%%%%%%%%%%%%%%%%%%%%%

%%%%%%%%%%%%%%%%%%%%%%%%%%%%%%%%%%%%%%%%%%%%%%%%%%%%%%%
\begin{figure*}
\begin{center}
\includegraphics[width=5.5cm]{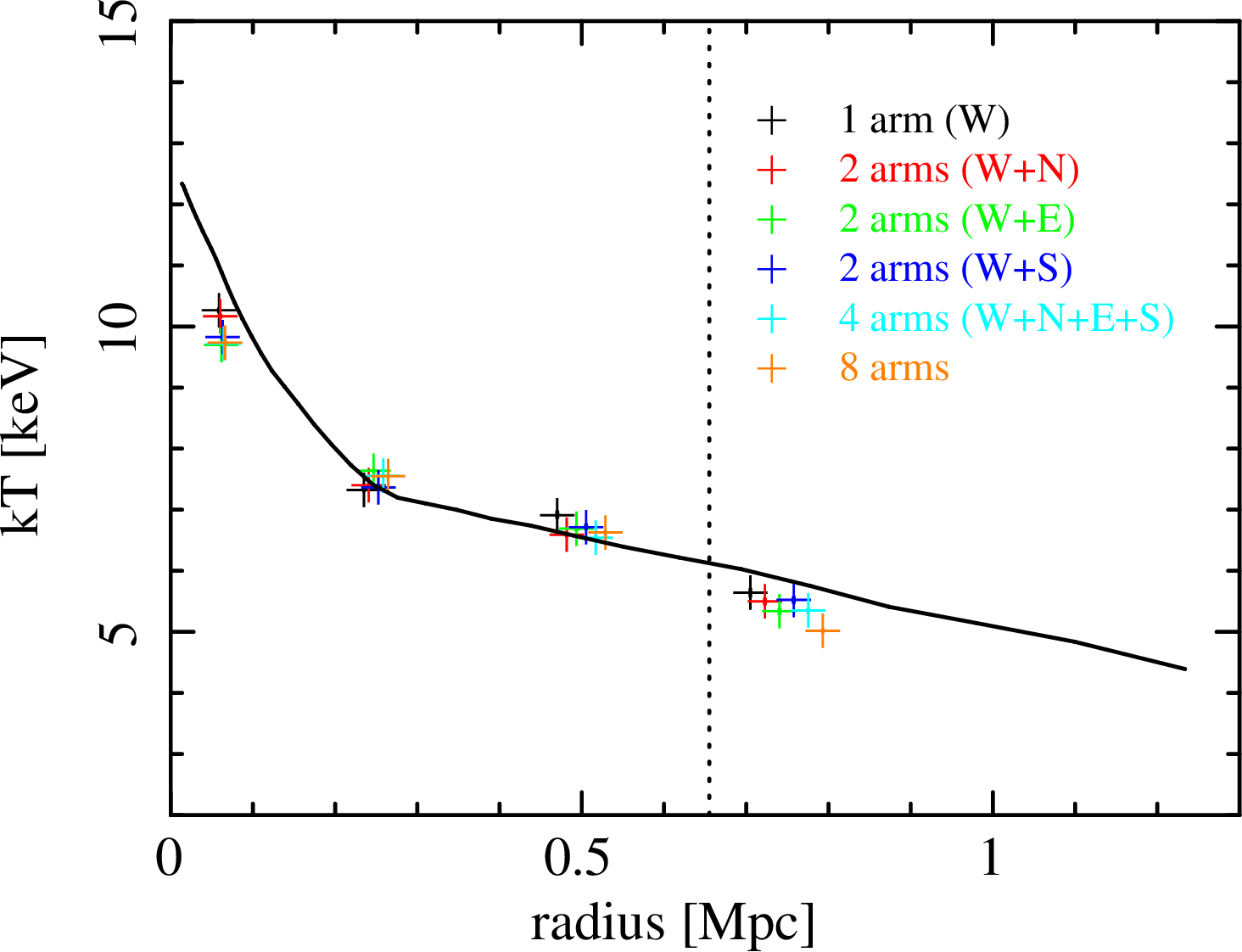}
\includegraphics[width=5.5cm]{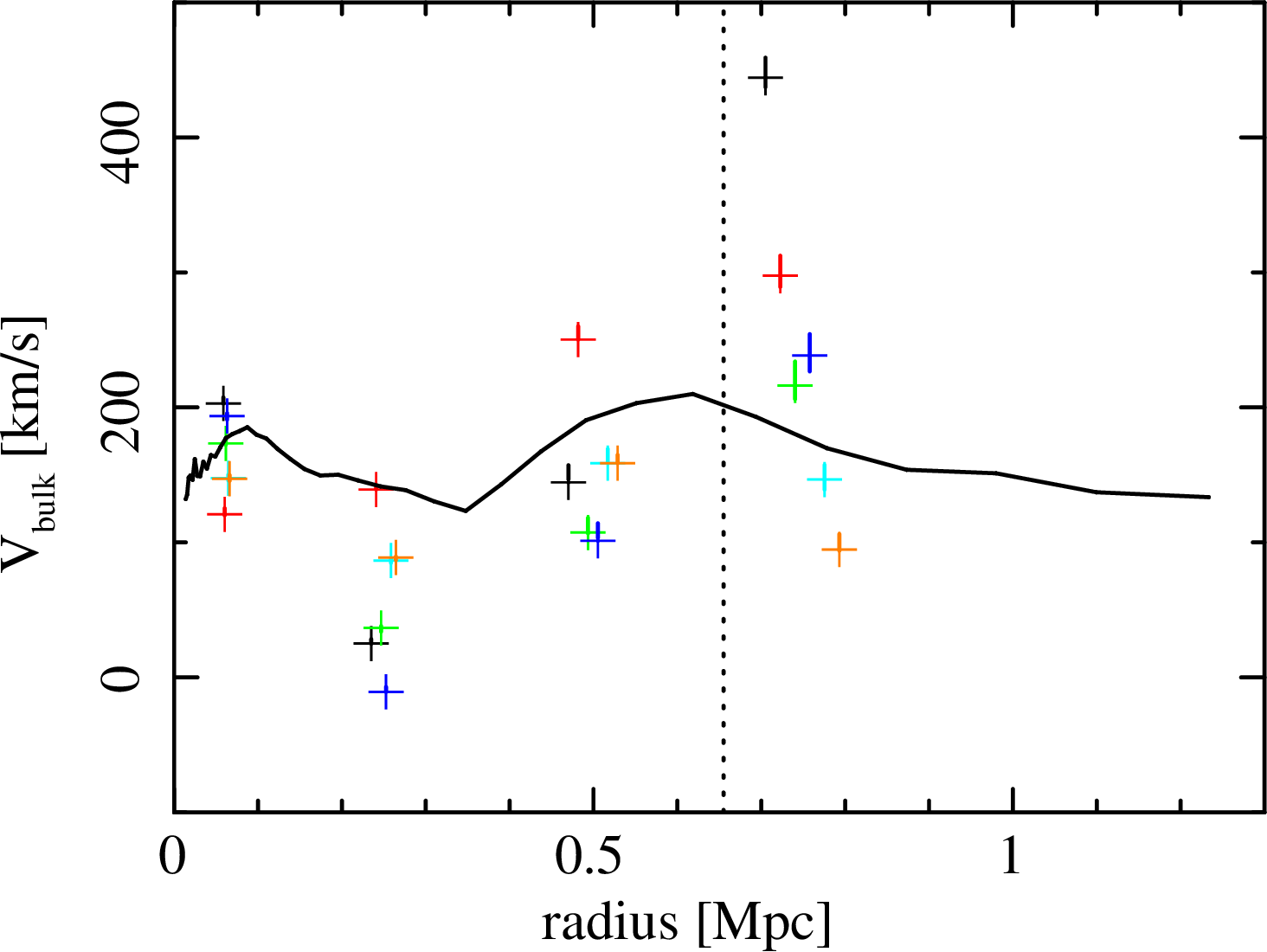}
\includegraphics[width=5.5cm]{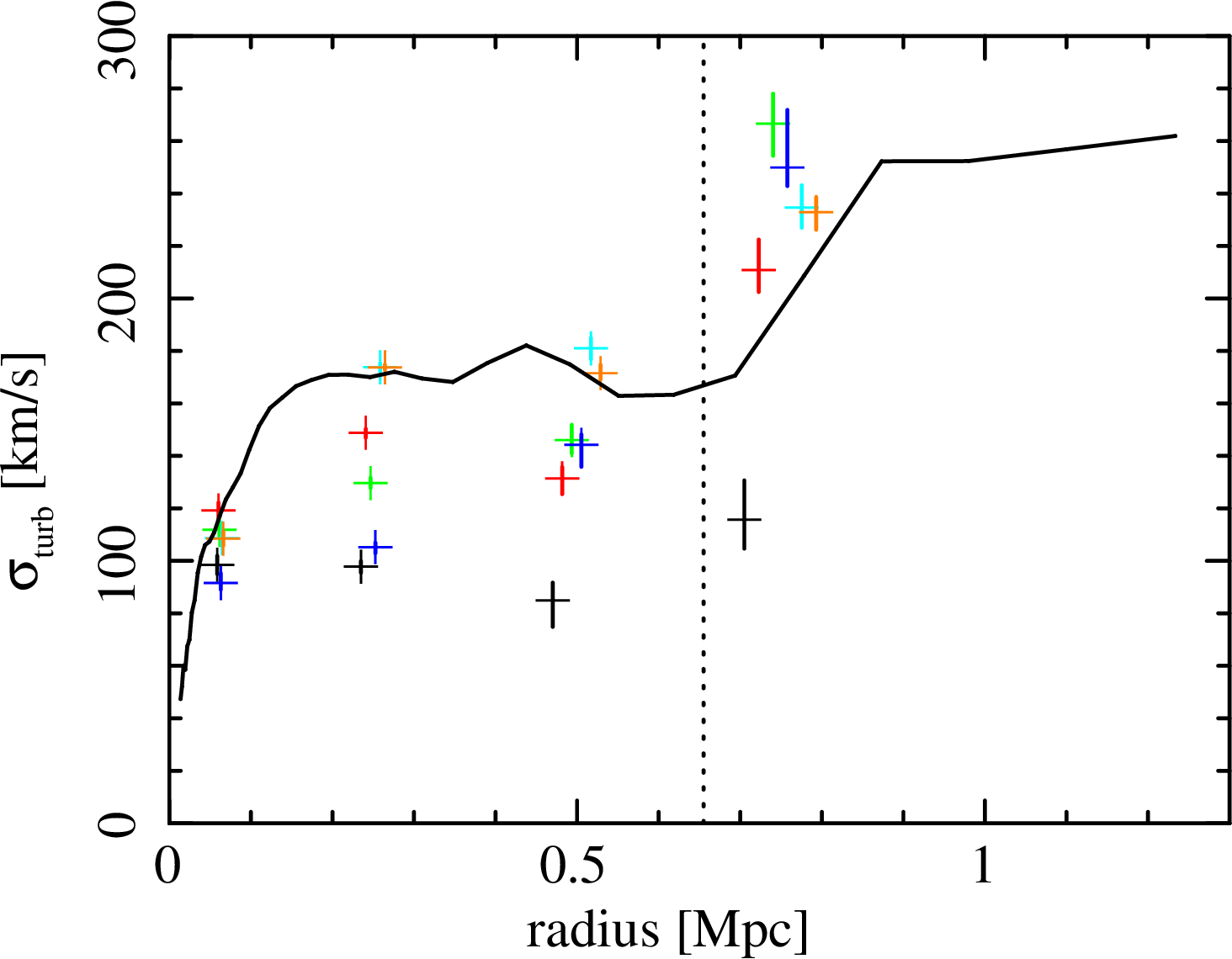}
\end{center}
\caption{
Effects of sample size. Results of deprojection analysis for spectra extracted from one arm (W), two arms (W+N, W+E, W+S), four arms (W+N+E+S), and eight arms are plotted. Deprojected temperature (top panels), bulk velocity (middle panels), and turbulent velocity (bottom panels) of the relaxed cluster CL104 viewed along the x projection (crosses). The Cash statistic was used to fit the 0.3--10~keV spectra. The circles indicate deprojected values. The solid lines indicate mass-weighted, spherically averaged values computed directly from simulation, and the dotted line indicates $r_{2500}$ of the cluster.   
\label{fig:deproj_box2_CL104}
}
%\vspace{5mm}
\end{figure*}
%%%%%%%%%%%%%%%%%%%%%%%%%%%%%%%%%%%%%%%%%%%%%%%%%%%%%%%

%-------------------------------------------------%
\subsection{Velocity profiles}
%-------------------------------------------------%

%-------------------------------------------------%
\subsubsection{Deprojection analysis}
%-------------------------------------------------%

One of the key science goals of XARM is to measure non-thermal pressure profile due to bulk and turbulent gas motions and their impact on the hydrostatic mass.  The non-thermal pressure profile is determined from spherically averaged density and velocity profiles. To study how well we can recover three-dimensional temperature and velocity profiles from the XARM spectra, we extract $0.3-10$~keV spectra from annular rings with $3\arcmin$ width (top left-hand panel in figure~\ref{fig:region}) and fitted them with the BAPEC model. 

Figure~\ref{fig:deproj_circle} show the projected and deprojected quantities for the $x$, $y$, $z$ projections for a relaxed cluster CL104. We compare the quantities measured from the mock against the true values measured directly from the simulation. 
The difference between the two is defined as $\Delta X = X_{\rm mock}/X_{\rm true} - 1$. 
Furthermore, we find that the deprojected values are in agreement with the mass-weighted, spherically averaged profiles measured directly from the simulation: $|\Delta T| <8\%$,  $|\Delta \sigma_{\rm turb}| < 28\%$ and $|\Delta V_{\rm bulk}| < 50\%$ at $r<r_{2500c}$. At larger radii, the photon counts in the \FeXXV~K$\alpha$ line drop below $200$, leading to larger differences with $\Delta T = 12\%$,  $\Delta \sigma_{\rm turb} = -4\%$ and $|\Delta V_{\rm bulk}| = 1.24$. 

%%%%%%%%%%%%%%%%%%%%%%%%%%%%%%%%%%%%%%%%%%%%%%%%%%%%%%%
\begin{figure*}
\begin{center}
\includegraphics[width=5.5cm]{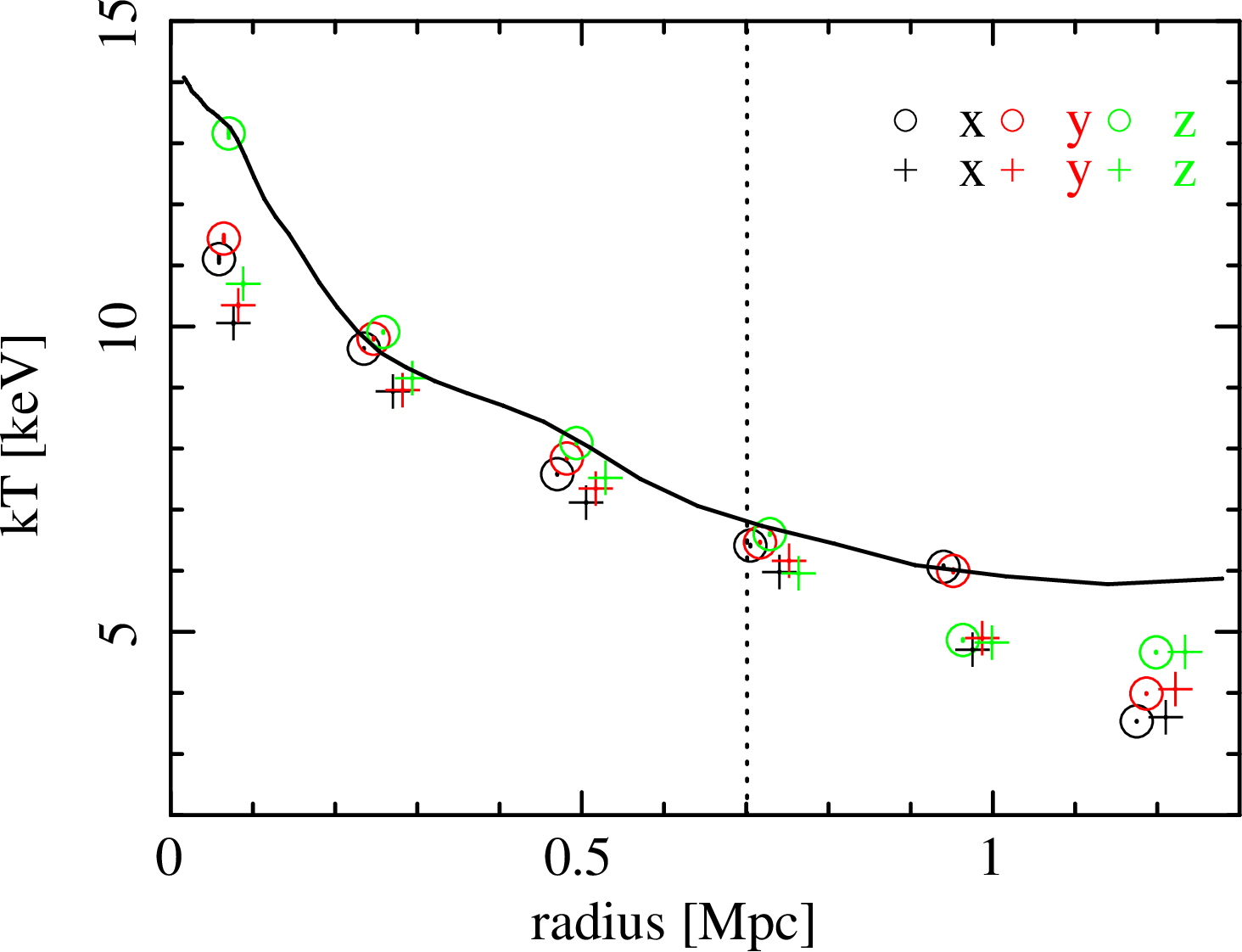}
\includegraphics[width=5.5cm]{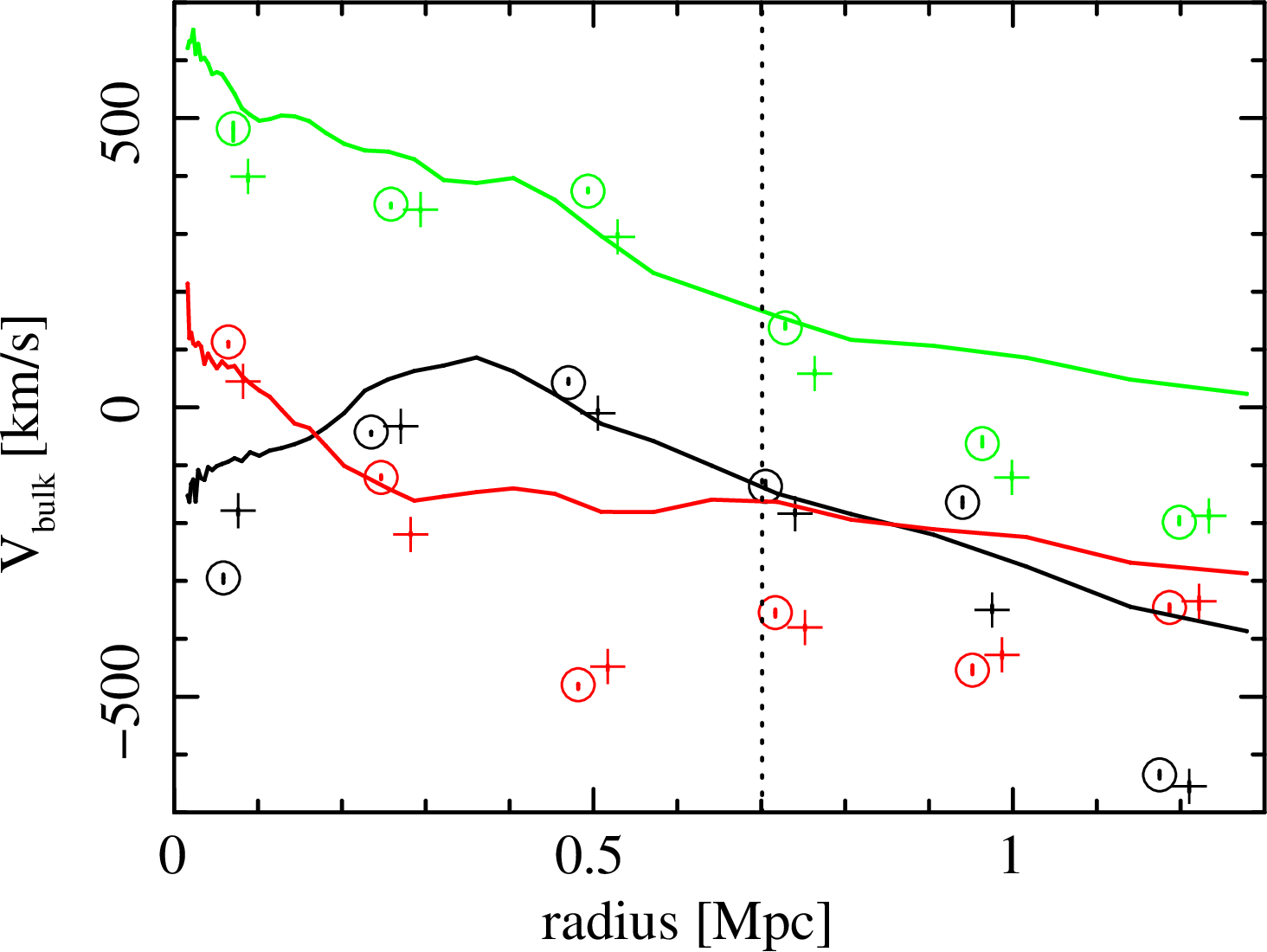}
\includegraphics[width=5.5cm]{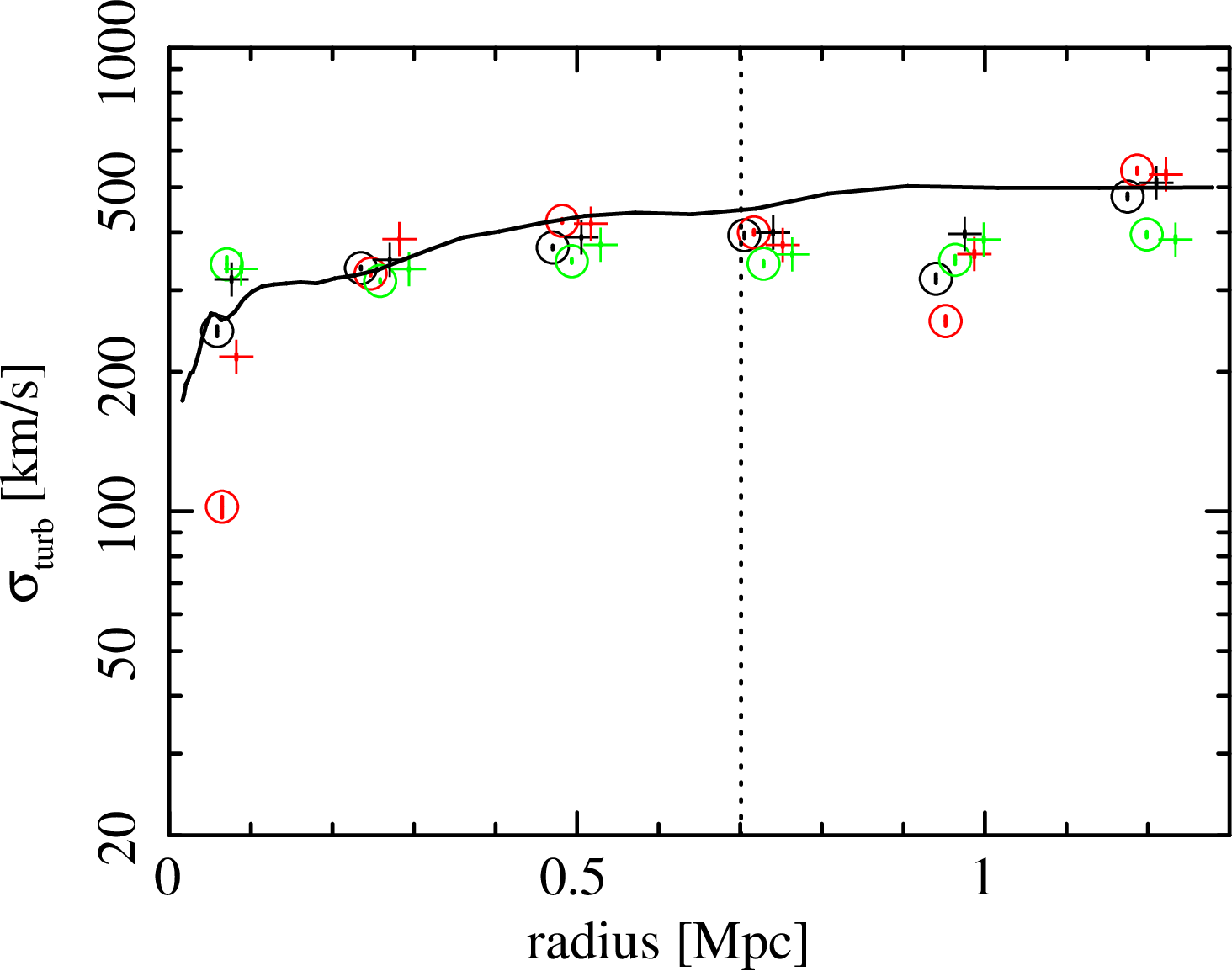}
\includegraphics[width=5.5cm]{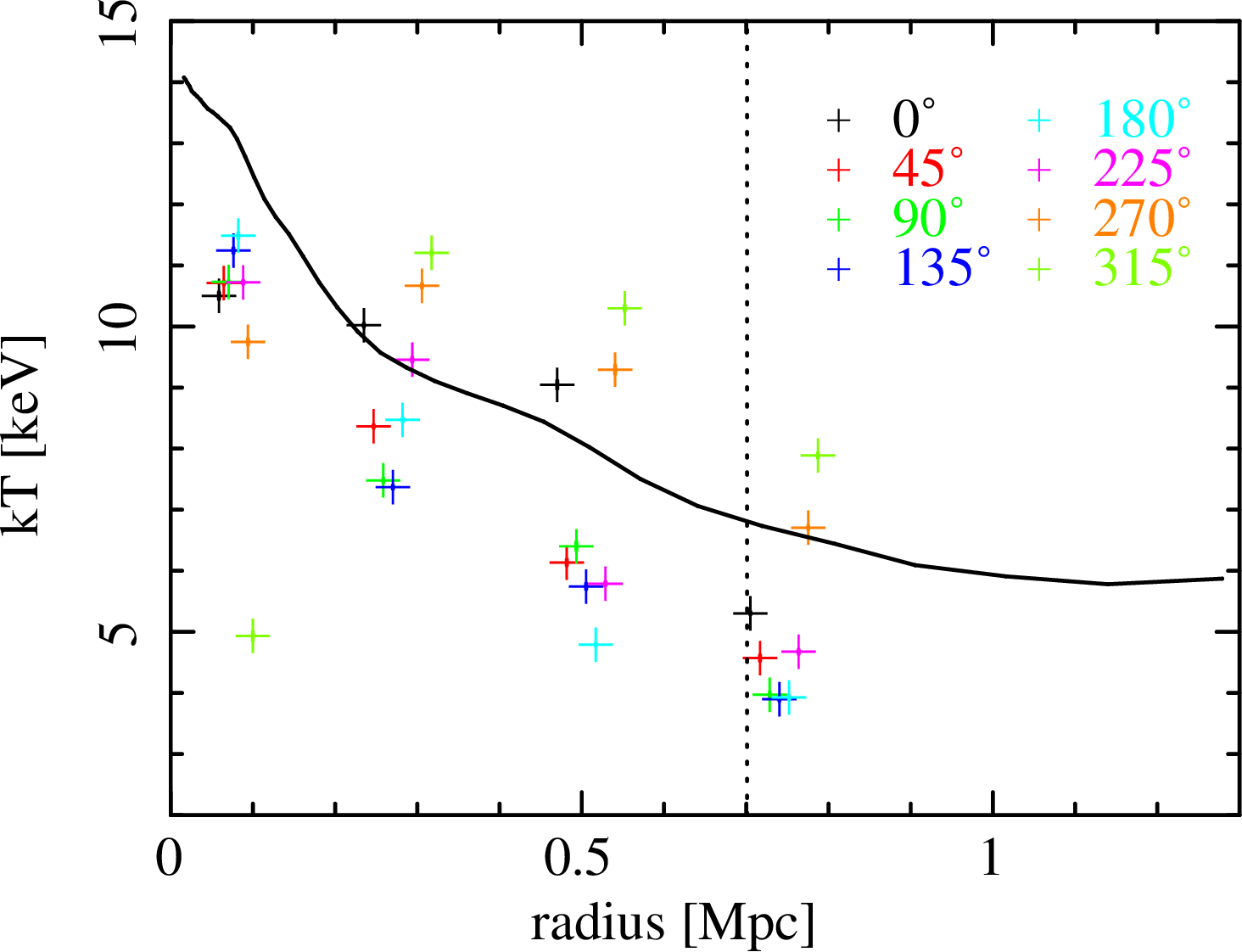}
\includegraphics[width=5.5cm]{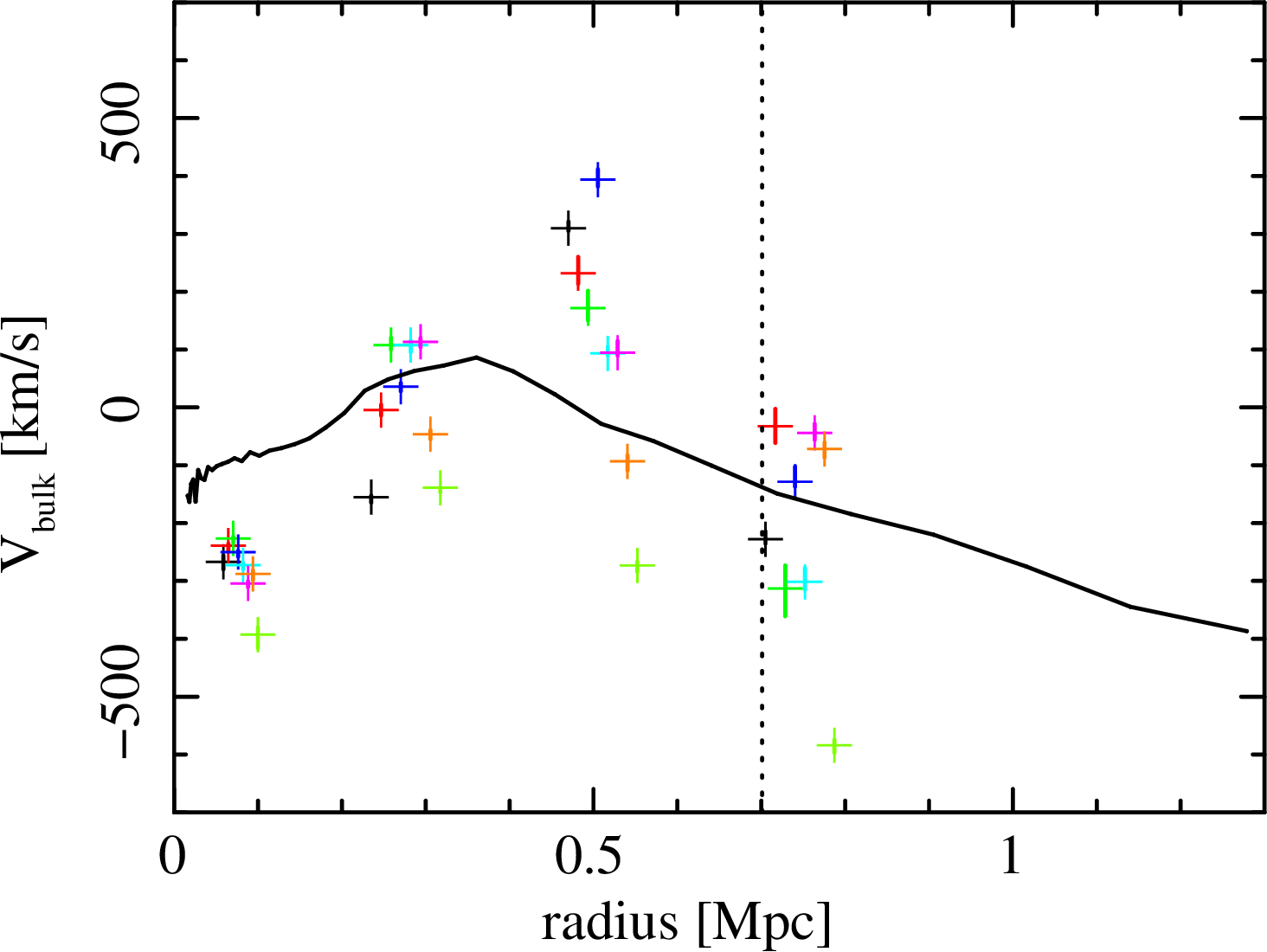}
\includegraphics[width=5.5cm]{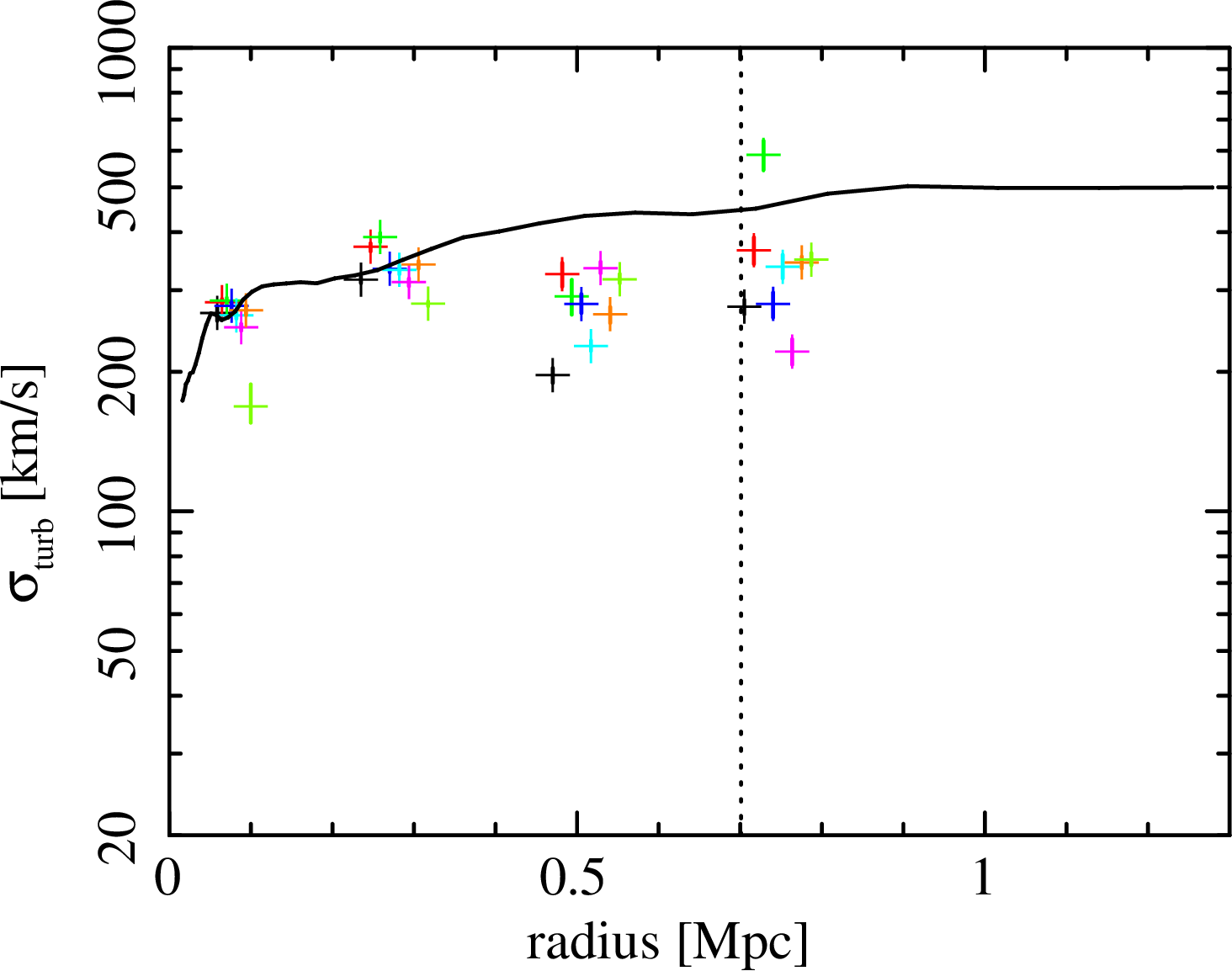}
\end{center}
\caption{Same as figure~\ref{fig:deproj_circle} for the merging cluster CL101. 
\label{fig:deproj_circle_CL101}}
%\vspace{5mm}
\end{figure*}
%%%%%%%%%%%%%%%%%%%%%%%%%%%%%%%%%%%%%%%%%%%%%%%%%%%%%%%

%-------------------------------------------------%
\subsubsection{Azimuthal variations}
%-------------------------------------------------%
\label{sec:azimuthal}

Since the field-of-view (FOV) of Resolve is small, we need to perform mosaic observations with multiple pointings to cover the spatially extended regions at large cluster-centric radii. However, since accurate velocity measurements require fairly deep exposures ($\gtrsim 100$~ks at $r\approx r_{2500}$), it would be costly to perform mosaic observations. In practice, observations at the large cluster-centric radii will be limited a few directions. Since the cluster gas is not perfectly spherically symmetric and its velocity distribution is generally anisotropic, we expect that the velocity profile measured along any given direction is not representative of the azimuthally averaged velocity profile in an annular ring.
 
In order to investigate the level of the azimuthal variations in the recovered velocity profiles, we extract spectra from four grids at $r\lesssim r_{2500}$ in eight azimuthal directions: $\phi = $ 0\degree, 45\degree, 90\degree, 135\degree, 180\degree, 225\degree, 270\degree, and 315\degree from the horizontal, corresponding to W, NW, N, NE, E, SE, S, and SW (top right-hand panel in figure~\ref{fig:region}). The deprojected temperature and velocity profiles of the eight azimuthal directions in the $x$ projection is plotted in figure~\ref{fig:deproj_circle} for CL104 in the left- and right-hand panels, respectively. 
For a given quantity $X$,  the azimuthal difference is defined as the $X$ measured in a single arm compared to the spherically averaged value directly measured from the simulation: $\Delta X_{\rm arm} \equiv X_{\rm arm}/X_{\rm true}-1$. 
Even in the relaxed cluster CL104, we find azimuthal variation in the fitted temperature, bulk and turbulent velocities. For example, the azimuthal variation in the projected temperature increases with radius, varying from $\Delta T_{\rm arm}(\min,\max)= (-0.02,0.11)$ at $r = 0.18 r_{2500}$ to $\Delta T_{\rm arm}(\min,\max) = (-0.09,0.24)$ at $r = 0.9 r_{2500}$. The temperature variation in the outer bin is driven by the infalling cold substructure in the NW direction. The same substructure is also responsible for the large difference between the bulk and turbulent velocities.  At $r = 0.9 r_{2500}$, the minimum and maximum azimuthal variations are $\Delta \sigma_{\rm turb, arm}(\min,\max)  = (-0.69, 0.32)$ for the turbulent velocity and $\Delta V_{\rm bulk, arm}(\min,\max)  = (-1.4,0.84)$ for the bulk velocity, respectively.

These results indicate that the velocity profile measured along any single arm is not representative of the spherically averaged profile. Specifically, the deprojected turbulent velocities along any single direction systematically under-estimate the spherically averaged turbulent velocity measured directly from the simulation (shown in the bottom right-hand panel of figure~\ref{fig:deproj_circle}), because the simple spherical analysis fails to capture the azimuthal variations in the bulk velocities (shown the bottom middle panel of figure~\ref{fig:deproj_circle}). Such variations in the bulk velocities are expected, because mergers and filamentary accretions occur anisotropically along the axes of mergers and filaments, respectively. 

In general, decomposition of gas motions into {\em turbulent} and {\em bulk} components, inferred from broadening and shifting of X-ray emission lines respectively, depends on the size and geometry of the regions where the X-ray spectra were extracted. To study the dependence of velocity recovery on the size of the regions, we co-added spectra at the same radii in different pointing directions (e.g., W and N) and performed the deprojection analysis.  Figure~\ref{fig:deproj_box2_CL104} shows the resulting recovered temperature, bulk, and turbulent velocity when the spectra in two, four or eight arms are combined. Compared to the results of a single arm, the turbulent/bulk velocity converges to the spherically averaged values when we combine spectra from more arms. Sampling more multiple directions leads more robust estimates of the spherically-averaged velocity profiles. Having eight arms leads to better recovery of the velocity profiles, with $\Delta \sigma_{\rm turb}\lesssim 23\%$ and $\Delta V_{\rm bulk} \lesssim 55\%$.

%%-------------------------------------------------%
\subsubsection{Dependence on dynamical states}
\label{subsubsec:dynamicalstate}
%%-------------------------------------------------%

Next, we repeat the same deprojection and azimuthal analyses on the merging cluster CL101. The bottom left- and right-hand panels of figure~\ref{fig:region} show the regions selected for these analyses, respectively. The upper panel in figure~\ref{fig:deproj_circle_CL101} shows its deprojected profiles for the temperature, bulk and turbulent velocities in annular bins and in the eight azimuthal directions, compared with the simulation values.  
The difference between the annular mock XARM values and the true simulation values for temperature and turbulent velocities are larger than the relaxed cluster, with $|\Delta T| <39\%$,  $|\Delta \sigma_{\rm turb}| < 45\%$ for all radial bins. The bulk velocities deviate significantly from the true values with $|\Delta V_{\rm bulk}| \sim 2.5$. At the merger shock (whose location is indicated in Fig.~\ref{fig:region}), $|\Delta V_{\rm bulk}|$ can reach $\sim 20$.  At $r=0.9 r_{2500}$, the azimuthal variation in temperature is larger than that of the relaxed cluster, with $\Delta T_{\rm arm}(\min, \max) = (-0.34,0.39)$. At the same radius, the variations in turbulent and bulk velocities are slightly larger than those of the relaxed cluster, with $\Delta \sigma_{\rm turb, arm}(\min,\max) = (-0.52,-0.22)$ and $\Delta V_{\rm bulk, arm}(\min,\max)  = (-5.5, 3.1)$.

%%%%%%%%%%%%%%%%%%%%%%%%%%%%%%%%%%%%%%%%%%%%%%%%%%%%%%%
\begin{figure}
\begin{center}
\includegraphics[width=7.5cm]{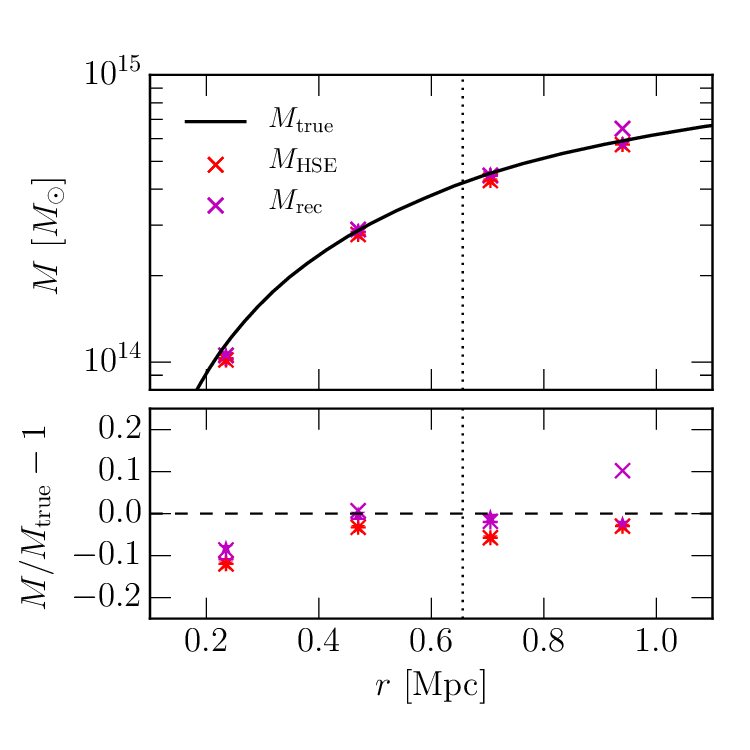}

\includegraphics[width=7.5cm]{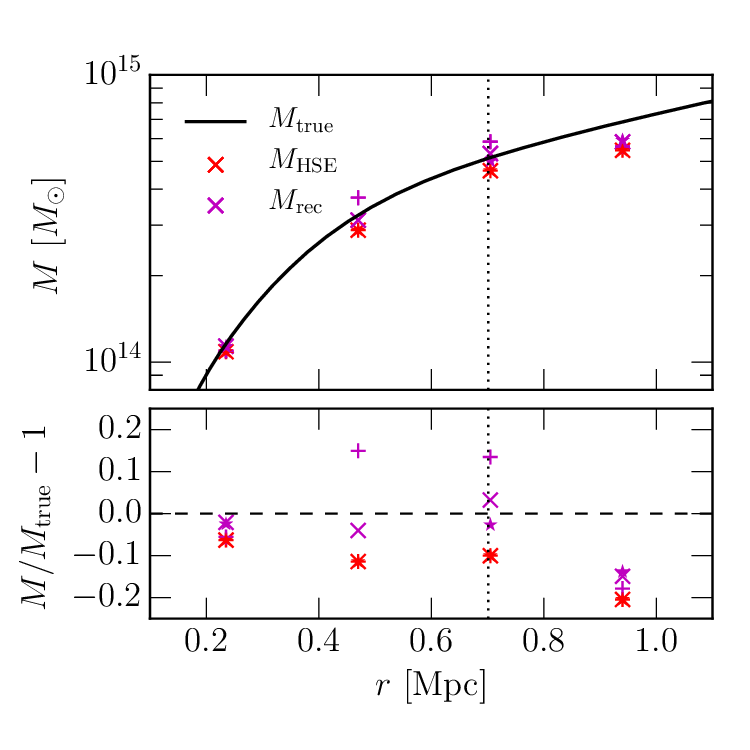}
\end{center}
\caption{
Recovery of the hydrostatic mass bias in the relaxed cluster CL104 (top panel) and the merging cluster CL101 (bottom panel).
Upper panels: The black line represents the true mass profile measured directly from simulation, the red points represent the hydrostatic mass $M_{\rm HSE}$ measured from simulation, and the magenta points represent the recovered mass profile, $M_{\rm rec} = M_{\rm HSE}+M_{\rm rand}+M_{\rm rot}$, using the velocity measurements from XARM mocks. Lower panels: Profiles of the hydrostatic mass bias, $M_{\rm HSE}/M_{\rm true}-1$ (red points) and the recovered mass, $M_{\rm rec}/M_{\rm true}-1$ (magenta points).  Different symbols indicate the recovered mass $M_{\rm rec}$ obtained from the XARM mock maps viewed along three orthogonal projections of each cluster. The vertical dotted lines indicate $r_{2500c}$ for each cluster. 
\label{fig:mpro}
}
\end{figure}
%%%%%%%%%%%%%%%%%%%%%%%%%%%%%%%%%%%%%%%%%%%%%%%%%%%%%%%

%%%%%%%%%%%%%%%%%%%%%%%%%%%%%%%%%%%%%%%%%%%%%%%%%%%%%%%
\begin{figure}
\begin{center}
\includegraphics[width=7.5cm]{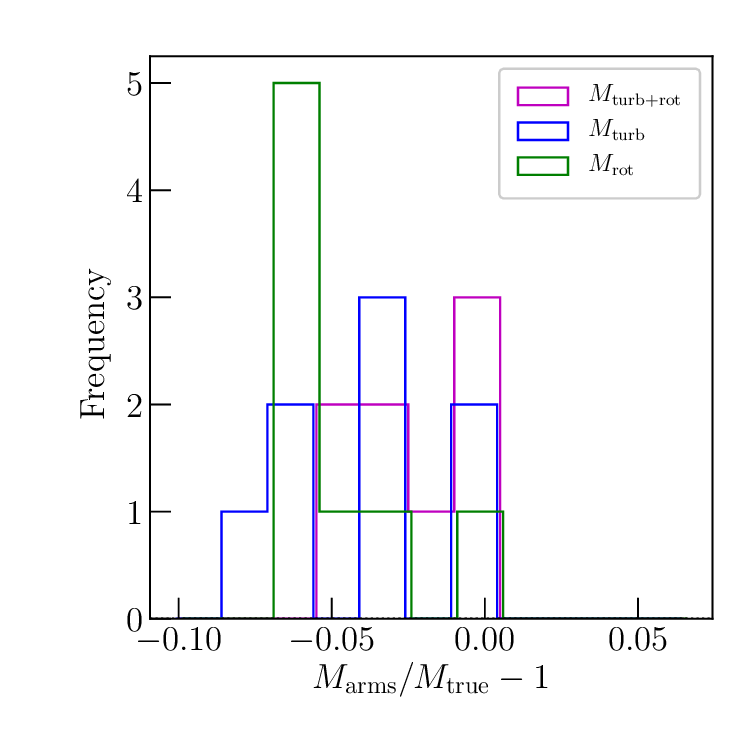}

\includegraphics[width=7.5cm]{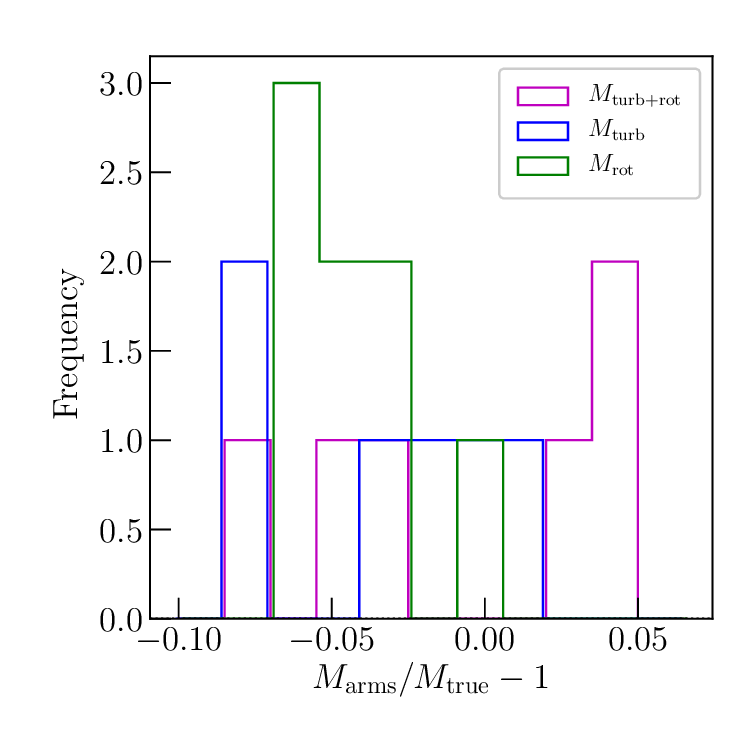}
\end{center}
\caption{
Histograms of the mass bias of the recovered mass computed from the deprojected bulk and turbulent velocity measurements along individual azimuthal arms $M_{\rm arm}/M_{\rm true}-1$, at $r=0.5$~Mpc (corresponding to $r_{2500c}$),  for the relaxed cluster CL104 (top panel) and the merging cluster CL101 (bottom panel). The magenta lines represent the recovered mass computed from combined turbulent and bulk motion measurements, while the the blue and green lines represent the mass computed using only turbulent and bulk motions, respectively. 
\label{fig:mbiashist}
}
\end{figure}
%%%%%%%%%%%%%%%%%%%%%%%%%%%%%%%%%%%%%%%%%%%%%%%%%%%%%%%

%-------------------------------------------------%
\subsection{Hydrostatic mass bias}
%-------------------------------------------------%
\label{sec:mass}

One of the primary sources of astrophysical uncertainties in cluster cosmology lies in hydrostatic mass bias due to non-thermal pressure support provided by internal gas motions in the ICM. 
Assuming spherical symmetry, the total cluster mass within some radius $r$ can be expressed as the following: 
\begin{equation}
M (<r) \cong M_{\rm HSE} + M_{\rm turb} + M_{\rm rot} + M_{\rm accel}. 
\end{equation}
where $M_{\rm HSE}$ is the hydrostatic mass, $M_{\rm rand}$ is the mass term associated with turbulent gas motions, $M_{\rm rot}$ is the term associated with the bulk rotational motions, and $M_{\rm accel}$ is the term associated with acceleration (or deceleration) of gas \citep[see][for details]{lauetal13}. Note that other mass terms are negligible ($\lesssim$1\%) compared to $M_{\rm turb}$, $M_{\rm rot}$ \citep{lauetal09} and $M_{\rm accel}$ \citep{nelsonetal14a}. The hydrostatic mass is given by
\begin{equation}
M_{\rm HSE} =  \frac{-r k T}{G\mu m_p}\left(\frac{\partial \log \rho} {\partial \log  r} + \frac{\partial \log T }{\partial \log  r}\right),
\end{equation} 
where $T$ and $\rho$ are the mass-weighted gas temperature and gas density, respectively. Note that the hydrostatic mass depends on the gradients of gas temperature and gas density, which may not be well determined with XARM owing to its relatively poor spatial resolution of  $1\arcmin3$.  The mass terms due to gas velocity $M_{\rm turb}$ and $M_{\rm rot}$, however, are less sensitive to resolution because both the average bulk and turbulent velocities do not change significantly with cluster-centric radius, and thus can be determined rather well with XARM: 
\begin{eqnarray}
M_{\rm turb}&=& \frac{-r^2}{G\rho}\frac{\partial\rho \sigma_{r}^{2}}{\partial r} - \frac{r}{G}\left(2\sigma_{r}^2-\sigma_{t}^2\right), \\
M_{\rm rot} &=& \frac{r}{G}\aav{v_{\rm rot}}^2,
\label{eqn:sph_mass}
\end{eqnarray}
where $\sigma_{r}^2$ and $\sigma_{t}^2$ are the mass-weighted velocity dispersion for the radial and tangential component, and $\aav{v_{\rm rot}}$ is the mass-weighted mean rotational velocity. 

The computation of these mass terms requires three-dimensional velocity fields that are not directly accessible to XARM observations, which measure only two components: the line-of-sight turbulent velocity ($\sigma_{\rm LOS}$) and the line-of-sight bulk velocity ($v_{\rm LOS}$) through broadening and shifting of X-ray emission lines, respectively. In order to estimate $M_{\rm turb}$ from observational data, we must assume the gas velocity dispersion is isotropic (i.e., the measured line-of-sight turbulent velocity is assumed to be equal to the isotropic one-dimensional velocity dispersion: $\sigma_{\rm LOS}^2 = \sigma_{\rm 1D}^2 = \sigma_r^2 = \sigma_t^2/2$).  Cosmological hydrodynamical simulations indicate that the gas velocity dispersion is approximately isotropic around our radius of interest ($r_{2500} \leq r \leq r_{500}$) on {\em average}, although there are significant cluster-to-cluster variations where the velocity anisotropy parameters ranging between $\beta=1-\sigma_t^2/(2\sigma_r^2) \approx \pm 0.5$ \citep[see e.g., figure~1 in][] {lauetal09}. For $M_{\rm rot}$, we estimate the rotational velocity $\aav{v_{\rm rot}}$ as follows. For each pair of opposite azimuthal arms, we measure the difference $\Delta v_{\rm LOS}$ of their line-of-sight bulk velocities ($v_{\rm LOS}$) . For example, for the NW-SE pair of arms, we measure $\Delta v_{\rm LOS} = |v_{\rm LOS, NW}-v_{\rm LOS,SE}|$. Among all pairs, we then take the largest $\Delta v_{\rm LOS}$ as the rotational velocity $\aav{v_{\rm rot}}$. Note that the rotational velocity estimated in this way is subjected to projection effect, as we have assumed that the rotation axis lies in the plane of the sky. This can lead to underestimate in the rotational velocity and therefore $M_{\rm rot}$. Despite these assumptions, the recovered mass is still good to $\sim 5\%$ as we show below.

The upper panels of figure~\ref{fig:mpro} show the mass profiles of the two clusters: CL104 (top panel) and CL101 (bottom panel). The black lines represent the true mass profile measured directly from simulation, the red lines represent the hydrostatic mass $M_{\rm HSE}$ measured from simulation, and the magenta points represent the recovered mass $M_{\rm rec} = M_{\rm HSE}+M_{\rm turb}+M_{\rm rot}$, with the latter two terms computed with velocity profile measurements from XARM mocks. Different symbols indicate the recovered mass $M_{\rm rec}$ obtained from the XARM mock maps viewed along three orthogonal projections of each cluster. The lower panels of figure~\ref{fig:mpro} show the profiles of the hydrostatic mass bias and the recovered mass, relative to the true cluster mass, for the same clusters. The hydrostatic mass bias is shown in red, and the bias of the recovered masses $M_{\rm rec} = M_{\rm HSE}+M_{\rm turb}+M_{\rm rot}$ is shown in magenta.  

In the lower panels, we demonstrate that XARM cluster observations can be used to recover the total cluster mass at $r\gtrsim r_{2500}$ at the level of about $5\%$ for dynamically relaxed clusters, provided that we measure both $M_{\rm turb}$ (turbulent motion) and $M_{\rm rot}$ (bulk rotational motion) terms from the line-of-sight velocity measurements. For example, for the relaxed cluster CL104, the remaining bias in the corrected mass, $M_{\rm rec}/M_{\rm true}-1$, varies from $+1\%$ to $-4\%$ in the outer three bins, compared to $-3\%$ to $7\%$ for the hydrostatic mass.  For the unrelaxed cluster CL101, the improvement in mass measurements is more evident; i.e., the remaining bias in the corrected mass, $M_{\rm rec}/M_{\rm true}-1$, improves from $-9\%$ to $+3\%$ in the inner bins. However, the mass recovery is not as accurate in the outer bins where the cluster gas is more out of equilibrium due to mergers and accretion. 

As shown in sub-subsection~\ref{sec:azimuthal}, the velocity measurements along individual azimuthal arms are not representative of spherically averaged velocity profiles, which could introduce biases in the recovered mass. Here, we demonstrate that this mass bias due to azimuthal variations in velocity can be mitigated provided that we account for both bulk and turbulent gas motions from line shifts and broadening, respectively (see the appendix for discussion). In figure~\ref{fig:mbiashist}, we show the histograms of the recovered mass computed from the deprojected bulk and turbulent velocities in each azimuthal arm $M_{\rm arm}$, compared to true cluster mass at $r=0.5$~Mpc. 
Note that we assumed that the rotational velocity is equal to the deprojected bulk velocity for each arm. For the relaxed cluster CL104, the recovered mass for a single arm deviates from the true mass from $-8\%$ to $0.1\%$ with only turbulent motions and  $-6\%$ to $-1\%$ with only rotational motions.  The mass recovery improves when we include both turbulent and bulk motions, leading to smaller bias and narrower distribution in the bias: $-5\%$ to $0.2\%$. For the merging cluster CL101, the mass recovery from a single arm also improves if we consider both bulk and turbulent motions.  

%-------------------------------------------------%
\section{Discussion}
\label{sec:discussion}
%-------------------------------------------------%

There are several effects that are missing from the present mock XARM analysis. We briefly discuss their effects on our results.

\vspace{2mm} \noindent
{\em Point Spread Function}: A systematic effect may arise from scattering of photons by a relatively wide Point Spread Function (PSF) of the mirror onboard XARM. Since XARM will have the same Soft X-ray telescope as Hitomi, we use the PSF of Hitomi to estimate the effect of PSF in XARM. The half power diameter of Hitomi is 1.3$\arcmin$ \citep{takahashietal16}. Strongly peaked X-ray emission at the center of relaxed cluster may contaminate the measurement of turbulent velocity at fainter, outer regions. Abundance peak in cool core clusters will result in stronger contamination. In the case of A1795, the fraction of scattered photons in a SXS field of view placed at the scale radius $r_{2500}$ from adjacent regions is estimated to be 25\% based on SXS simulations assuming the azimuthally-averaged PSF model. This effect can be corrected by simultaneously analyzing a series of pointings along the radius from the center and solving intrinsic distributions of the line shifts and width \citep{kitayamaetal14}.

\vspace{2mm} 
\noindent {\em Baryonic physics}: The results presented in this paper are based on non-radiative simulations that do not include baryonic effects such as radiative cooling, star formation, and AGN feedback. These baryonic physics can alter the velocity structure of the ICM especially in cluster cores, while the velocity structure outside the cluster core ($r\gtrsim 0.15 r_{500}$) is relatively unaffected by the cluster core physics \citep{nagaietal13}. While AGN feedback in the form of bubble or jet alone can be too inefficient to drive the observed level of turbulent motions \citep{reynoldsetal15, bournesijacki17}, hydrodynamical cosmological simulations that include the effects of both cosmic accretion and AGN bubbles are capable of reproducing the level of large-scale velocity shear and Doppler broadening measured in the Perseus Cluster \citep{lauetal17}. Further theoretical works are needed to understand the mechanisms of driving ICM motions as well as the roles and relative importance of AGN bubbles and cosmic accretion in and beyond the cluster core.

\vspace{2mm} 
\noindent {\em Plasma Instabilities}: Our cluster simulations do not model magnetic field and plasma effects, which could amplify turbulent motions in the ICM outside the cluster core, e.g., through the magneto-thermal instability and increase the non-thermal pressure support by $\sim 50\%$ at $r_{2500}$ \citep{parrishetal12}. Magnetic fields can also suppress fluid instabilities perpendicular to the field lines, leading to suppressed mixing and less turbulence. XARM observation should be able to constrain the effects of magnetic fields and plasma instabilities with its gas motion measurements. 

\vspace{2mm} \noindent
{\em Resonant Scattering}: We did not include the effects of resonant scattering, e.g., in the  the 6.7~keV \FeXXV~K$\alpha$ line, which has been observed by Hitomi in the core of the Perseus cluster \citep{hitomi17b}.  Resonant scattering increases the optical depth and suppresses the strength of the emission line. The effect of resonant scattering can be important where the density, and therefore the optical depth to \FeXXV~K$\alpha$ line is high. Outside the central region, however the effect is expected to be small. In the presence of turbulent motions, resonant scattering is suppressed by Doppler shifting the line out of resonance. Adding the effect of resonant scattering can  improve the velocity recovery of XARM by providing additional constraints from the shape of the resonant line in the cluster core, but probably not in the outer regions where density is lower. 

%-------------------------------------------------%
\section{Conclusions}
\label{sec:conclusions}
%-------------------------------------------------%

Gas motions play an important role in determining the properties of the hot X-ray emitting ICM and in the constraint of cosmological parameters using X-ray and SZ effect observations of galaxy clusters.  
Despite their importance, observationally we know very little about the nature of gas motions in the ICM.  Hydrodynamical simulations are capable of predicting the nature of gas motions generated by large-scale cosmic accretion and mergers during the hierarchical build of galaxy clusters. 
The upcoming XARM mission will provide the first direct measurements of gas motions in galaxy clusters through Doppler shifting and broadening of X-ray emission lines,  allowing us test simulation prediction against observation. 
In this work, we investigate how well XARM can measure the bulk and turbulent gas motions in the ICM by analyzing mock XARM maps and spectra of simulated clusters extracted from hydrodynamical cosmological simulations. We summarize our main findings below:

\begin{enumerate}
\item XARM spectra extracted from annular bins give a good recovery of the thermodynamic and velocity profiles of the ICM for dynamically relaxed clusters. Specifically, deprojected temperature and turbulent velocity profiles from spectra extracted from the annular bins differ from the ``true'' mass-weighted, spherically averaged profiles computed directly from the simulations by less than $30\%$, provided that there are enough photons ($N>200$) to resolve the \FeXXV~K$\alpha$ line.  For dynamically disturbed clusters the agreement worsens to about $45\%$. 

\item We find significant azimuthal variations in the recovered turbulent velocities. The difference in the turbulent velocity in one azimuthal direction can deviate from the ``true'' mass-weighted spherically averaged value by a factor of $\sim 1.4$, even in dynamically relaxed simulated clusters. This indicates that the velocity structure of the ICM is richer and more complex than its thermodynamic counterpart.  Measurement of the turbulent velocity profile in any given azimuthal direction (or arm) is, therefore, not representative of the underlying mass-weighted, spherically averaged profile, which is required to correct for the hydrostatic mass bias. Since the field of view of XARM's spectrometer is small, mosaic observations are required to sample the annular regions at the large cluster-centric radii of nearby clusters. Our results indicate that combining spectra from eight arms leads to better recovery of the velocity profiles, with a difference of  less than about 30\% compared to the spherically averaged values computed directly from the simulation. 

\item We demonstrate that XARM observations should be able to recover the total mass at $r\lesssim r_{2500}$ at the level of $\sim 5\%$ for dynamically relaxed cluster by accounting for {\em both} bulk and turbulent gas motions through line-of-sight Doppler broadening and shift measurements with XARM, despite several simplifying assumptions (such as spherical symmetry for the ICM spatial distribution and isotropy of gas velocities) and spectroscopic measurements in a finite set of azimuthal arms. 
Our finding indicates that XARM can measure the hydrostatic mass bias associated with the internal gas motions, provide first direct test of the simulation predictions, and open up a possibility of correcting for the hydrostatic mass bias using real data.

\end{enumerate}

Although our analyses have focused on XARM, some of these results are broadly applicable to future X-ray missions, such as Athena\footnote{{http://www.the-athena-x-ray-observatory.eu/}} and Lynx\footnote{{https://wwwastro.msfc.nasa.gov/lynx/}}. The improved sensitivities and angular resolutions of the future missions should produce spatially-resolved velocity structure in the ICM \citep{biffietal13,fujitaetal05}, resolve the dissipation scales of ICM motions \citep{zuhoneetal16}, and probe the physical nature of the ICM plasma \citep{gasparichurazov13, gasparietal14} through the measurements of the ICM velocity power spectrum \citep{zhuravlevaetal12,zhuravlevaetal14}.  However, the future measurements will be increasingly limited by systematic uncertainties (associated e.g., with multiphase ICM and dynamical state of clusters) rather than statistical uncertainties. Thus, future work must address how to separate bulk and turbulent motions, how well different codes agree on the gas velocity field, and what are the main drivers of bulk and turbulence motions.

\begin{ack}
We acknowledge Miyu Masuyama for her contribution to the early stage of this work and Eugene Churazov, Kyoko Matsushita, Takaya Ohashi, Andrew Szymkowiak and Irina Zhuravleva for discussions and/or comments on the manuscript. This  was supported in part by JSPS, KAKENHI grant 2540023, 25247028, and 16K05295, NSF grant AST-1412768 \& AST-1009811, NASA ATP grant NNX11AE07G, NASA Chandra Theory grant GO213004B, the Research Corporation, and by the facilities and staff of the Yale Center for Research Computing.
\end{ack}

\bibliographystyle{aasjournal}
\bibliography{ref}

\appendix
\section*{Decomposition of bulk and turbulent gas motions and hydrostatic mass bias}\label{sec:app}
In \citet{lauetal13}, we showed that there are two methods for measuring hydrostatic mass bias from gas motions, and that they are mathematically equivalent. In this appendix, we show how this result can be applied to in the correction of hydrostatic mass bias from XARM measurements. 

First we give a brief review of the two methods for measuring hydrostatic mass bias . The first method, called the ``Summation method'', is based on Gauss's Law and Euler's Equation, where the cluster mass within some volume $V$ with surface ${\partial V}$. Assuming steady-state of gas (for simplicity), the mass estimate obtained using this ``Summation method'' is given by
\begin{equation}
M^S = \frac{-1}{4\pi G}\int_{\partial V} \left(\frac{\nabla P}{\rho} +(\mathbf{v}\cdot \nabla )\mathbf{v}\right) \cdot d\mathbf{S},
\label{eq:sum_mass}
\end{equation}
where $\mathbf{v}$ is the gas velocity on the infinitesimal surface element $d\mathbf{S}$, and $\rho$ and $P$ are the corresponding gas density and pressure. We can recover the true mass of the cluster if we measure $\mathbf{v}$, $\rho$ and $P$ of individual gas element and sum them up over the entire surface using Equation~(\ref{eq:sum_mass}). In reality, it is not possible to do so since observations are limited by finite spatial resolution of the instrument and we cannot measure the physical quantities of each individual gas element. In practice, we measure spatially averaged values above some resolution scale. To mimic real measurement, we can spatially average Equation~(\ref{eq:sum_mass}) over the surface ${\partial V}$ to arrive at a cluster mass estimate that is based on the spatial averages of gas velocity, density and pressure. The mass obtained using this ``Averaging method'' is given by
\begin{equation}
M^A = \frac{-r^2}{G\langle\rho\rangle} \nabla \langle P \rangle +\frac{-r^2}{G\langle{\rho}\rangle} \nabla \langle{\rho}\rangle \sigma^{2} +\frac{-r^2}{G}(\langle \mathbf{v} \rangle_{\rho}\ \cdot \nabla) \langle \mathbf{v} \rangle_{\rho},
\label{eq:avg_mass}
\end{equation}
where the angular bracket $\langle \cdots \rangle$ denotes spatial averaging, and $\langle \cdots \rangle_\rho$ denotes spatial averaging weighted by the gas density $\rho$ over a spherical shell with radius $r$. 
Also note that $M^A$ contains a density-weighted velocity dispersion tensor $\sigma^{2}_{ij} \equiv \langle (v_i- \langle v_i \rangle_\rho)(v_j- \langle v_j \rangle_\rho) \rangle_\rho$. This term represents turbulent motions of the gas elements internal to the resolution scale, which manifests as Doppler broadening in the emission lines. The bulk motion of the gas at the resolution scale, on the other hand, is represented by $\langle \mathbf{v} \rangle_{\rho}$, which can be measured as Doppler shifts in the emission lines. Thus, by accounting for {\em both} the internal turbulent motions and the bulk motions, from Doppler line broadening and shifts, respectively, we should be able to recover the cluster mass. 

The relative contributions of bulk and turbulent motions in $M^A$ components depend on the scale of spatial averaging.  As the averaging scale increases, the turbulent motion term increasingly dominates as the motions become increasingly more uncorrelated . The recovered mass $M^A$ from the sum of bulk and turbulent gas motions, however, is independent of the averaging scale.  As mass is an extensive quantity and not affected by spatial averaging, $M^A$ and $M^S$ should be identical regardless of the averaging scale. Since $M^S$ is independent of the averaging scale, hence $M^A$ is also independent of the averaging scale, and therefore independent of the relative contributions of the bulk and turbulent gas motions. 

\end{document}